\documentclass[aps,preprintnumbers,nofootinbib,pacs, twocolumn]{revtex4}
\usepackage{eurosym}
\usepackage{graphicx}
\usepackage{amsmath}
\usepackage{graphicx}

\def\be{\begin{equation}}
\def\ee{\end{equation}}
\def\bea{\begin{eqnarray}}
\def\eea{\end{eqnarray}}

\begin{document}


\title{Null fluid collapse in brane world models}

\author{Tiberiu  Harko}
\email{t.harko@ucl.ac.uk} \affiliation{Department of Mathematics, University College London,
Gower Street, London WC1E 6BT, United Kingdom}
\author{Matthew J. Lake}
\email{matthewj@if.nu.th} \affiliation{ThEP's Laboratory of Cosmology and Gravity \\ The Institute for Fundamental Study ``The Tah Poe Academia Institute", Naresuan University, Phitsanulok 65000, Thailand}
\affiliation{Thailand Center of Excellence in Physics, Ministry of Education, Bangkok 10400, Thailand}


\begin{abstract}
The brane world description of our universe entails a large extra dimension and a fundamental scale of gravity that may be lower than the Planck scale by several orders of magnitude. An interesting consequence of this scenario occurs in the nature of spherically-symmetric vacuum solutions to the brane gravitational field equations, which often have properties quite distinct from the standard black hole solutions of general relativity. In this paper, the spherically-symmetric collapse on the brane world of four types of null fluid, governed by the barotropic, polytropic, strange quark ``bag" model and Hagedorn equations of state, is investigated. In each case, we solve the approximate gravitational field equations, obtained in the high density limit, determine the equation which governs the formation of apparent horizons and investigate the conditions for the formation of naked singularities. Though, naively, one would expect the increased effective energy density on the brane to favor the formation of black holes over naked singularities, we find that, for the types of fluid considered, this is not the case. However, the black hole solutions differ substantially from their general-relativistic counterparts and brane world corrections often play a role analogous to charge in general relativity. As an astrophysical application of this work, the possibility that energy emission from a Hagedorn fluid collapsing to form a naked singularity may be a source of GRBs in the brane world is also considered.
\end{abstract}


\maketitle

\section{Introduction}

Fundamental results in superstring theory and M-theory suggest that our four-dimensional world is embedded into a higher-dimensional space-time. In particular, ten-dimensional $E_8 \otimes E_8$ heterotic superstring theory is obtained as the low-energy limit of eleven-dimensional supergravity, under the compactification scheme $M^{10}\times S_1 / Z_2$ \cite{Witten,Witten2}. In order for this result to be compatible with the basic observation that we appear to live in a large, four-dimensional universe, the remaining ten-dimensional space-time must be decomposed as the cross-product of a four-dimensional manifold, with three large spatial dimensions and one dimension of time, and a six-dimensional space-like manifold, compactified on a much smaller hierarchy of length-scales, $M^4 \times CY^6 \times S_1 / Z_2$, implying that our universe exists as a $(3+1)$-dimensional ``brane" embedded into a higher-dimensional bulk space. In this paradigm, standard model particles are described by open string modes, in which the string end points are confined to the brane world, while gravitons are described by closed strings which propagate freely within the bulk \cite{Polchinski}.\\
\indent
In light of these fundamental results, numerous string-inspired phenomenological models of higher-dimensional cosmology have been devised \cite{Gi99}-\cite{BaLoMePedaSa13}. Among the collection of brane world scenarios, the Randall-Sundrum type II model has the virtue of providing a new kind of compactification of gravity \cite{Randall,Randall2}. In this scenario, standard four-dimensional gravity is recovered as the low-energy limit of the theory, in which a $3$-brane of positive tension is embedded in a five-dimensional anti-de Sitter bulk space. The covariant formulation of the brane world model, given by Shiromizu, Maeda and Sasaki \cite{Shiromizu}, leads to a modification of the standard Friedmann equations on the brane. In this scenario, the dynamics of the early universe may be substantially altered by the presence of additional quadratic terms  in the energy density, resulting from the brane tension, as well as by the effects of nonzero components of the bulk Weyl tensor, which both give contributions to the effective, four-dimensional, energy-momentum tensor. For general reviews of brane world gravity and resulting cosmologies, see \cite{GaTa00,SaSh02,Mart04,Du05,Cl07,Wa06,Ta08,Ma07,MaKo10} .\\
\indent
Similarly, these effects imply a modification of the basic equations describing cosmological and astrophysical dynamics, which have been extensively studied in the literature \cite{all2}. Specifically, several classes of static, spherically-symmetric, solutions of the vacuum field equations on the brane have been obtained ~\cite{HaMa03,Ma04,Ha04,Ha05}. As a possible physical application of these solutions the behavior of the angular velocity, $v_{tg}$, of test particles in stable circular orbits has been considered~\cite{Ma04,Ha05,Ta07,RaKaDeUsRa08,Na09,Geetal11}. The general form of the solution, together with the specific values of the two, arbitrary, constants of integration, uniquely determines the rotational velocity of the particle. Interestingly, for a particular range of values in the parameter space, the angular velocity tends to a constant for large radial distances. This behavior is typical for massive particles (e.g. hydrogen clouds) orbiting the outer regions of galaxies, and is usually explained by postulating the existence of the dark matter. The exact galactic metric, dark radiation, dark pressure and lensing effect in the flat rotation curves region of the brane world scenario were determined in \cite{Ha05}. Brane world dark energy has been studied in \cite{SaSh03,NeVa06,Ma07*,Ko08}.\\
\indent
For standard, general-relativistic, spherical compact objects the exterior space-time is described by the Schwarzschild metric. However, in five-dimensional brane world models, the brane-tension corrections to the energy density, together with the Weyl stresses from bulk gravitons, imply that, on the brane, the exterior metric of a static star is no longer Schwarzschild \cite{Da00}. The presence of the Weyl stresses also means that the matching conditions do not have a unique solution on the brane. In general, a knowledge of the five-dimensional Weyl tensor is required as a minimum condition for uniqueness. Static, spherically-symmetric, exterior vacuum solutions to the brane world field equations were first studied by Dadhich et al. \cite{Da00} and Germani and Maartens \cite{GeMa01}. The first of these solutions, obtained in \cite{Da00}, has the mathematical form of the Reissner-Nordstrom solution, in which a tidal Weyl parameter plays the role of the electric charge in the general-relativistic solution and was obtained by imposing the null energy condition on the $3$-brane for a bulk having nonzero Weyl curvature. The exterior geometry can be matched to the interior solution, corresponding to a brane world star of constant density. A second exterior solution, which also matches a constant-density interior, was derived in \cite{GeMa01}.\\
\indent
Two further classes of spherically-symmetric vacuum solutions in the brane world model (with $g_{tt}\neq -1/g_{rr}$), parameterized by the Arnowitz-Deser-Misner mass and a post-Newtonian parameter, $\beta $, were found by Casadio, Fabri and Mazzacurati \cite{cfm02}.
Non-singular black hole solutions have also been considered in \cite{Da03}, by relaxing the condition of zero scalar curvature but retaining the null energy condition. The ``on brane'', four-dimensional, Gauss-Codazzi equations for an
arbitrary, static, spherically-symmetric star in the Randall-Sundrum type II model have been completely solved by Visser and Wiltshire \cite{Vi03}. The on-brane boundary may also be used to determine the full, five-dimensional, space-time geometry and this procedure can be generalized to solid objects such as planets. A method to extend asymptotically flat, static, spherically-symmetric brane world metrics into the bulk was proposed by Casadio and Mazzacurati \cite{Ca03}, where the exact integration of the field equations along the fourth spatial coordinate was done by using the multipole ($1/r$) expansion. These results suggest that the shape of the horizon for brane black holes is very likely to be a flat ``pancake'' for astrophysical sources.\\
\indent
In \cite{BMD03}, the  general solution to the trace condition of the four-dimensional Einstein equations for static, spherically-symmetric, configurations was used to construct a general class of black hole metrics containing one arbitrary function, $g_{tt}=A(r)$, which vanishes at some distance, $r=r_h>0$ (the horizon radius). Under certain reasonable restrictions black hole metrics were shown to exist, either with or without matter. Depending on the boundary conditions, these metrics can be asymptotically flat, or have any other prescribed asymptotic behavior.\\
\indent
The formation of compact objects, gravitational collapse and singularity formation in the brane world have been extensively studied in recent years, from a number of perspectives. On-brane solutions corresponding to brane world stars \cite{Ov08,deLe08,Ov09,Ov10B,CaOv13,OvLi13} and collapse solutions ending in both black holes \cite{ChaHaRe00,BiChDaDu02,MaMu05,FiRaWi06,CrGrKaMi06,ShWaSu06,HeRaSe07,GeKi07,GiRoSz08,PuKoHa08,Ka004,BaBaCh08,Gr09,Mo09,Ka09,AlTa09,Ka09*,Ov10A,TaTa11,FiWi11,AnKu11,AmEi12,RaSeSh12,MaDa12,AkHaMu12} and naked singularities \cite{LoZe00,BrDa01} have been found, though the latter seem to have been comparatively neglected in the literature. Among the black hole solutions obtained, a class of topologically charged black holes exists in which bulk effects play a non-trivial role \cite{ShWa09,LaGrLo13} and bulk black hole solutions which effectively localize to the brane have been studied in \cite{KaPaZu13}. Certain aspects of the causal structure of on-brane and bulk black hole solutions have been studied by analyzing their time-like geodesics \cite{Yo01,ZhChWa11}  though, again, comparatively little work has been done in this area. For a review of the black hole properties and of lensing effects in brane world models see \cite{MaMu05}. Further work on lensing in brane worlds is presented in \cite{Ei05,PaKa08,MuMa09,Bi10,EiSe13} and current observational constraints on the model parameters from classical, solar-system, tests are contained in \cite{BoHaLo08,Bo10}.\\
\indent
Investigating the final fate of the collapse of an initially regular distribution of matter, in the framework of four-dimensional general relativity, is also one of the most active fields in contemporary gravitational research. One would like to know whether, and under what initial conditions, gravitational collapse results in black hole formation. One would also like to know if there are physical solutions that lead to naked singularities. If found, such solutions would be counterexamples of the cosmic censorship hypothesis, which states that curvature singularities in asymptotically flat space-times are always shrouded by event horizons.\\
\indent
This idea, also known as the cosmic censorship conjecture, was first proposed by Penrose \cite{Pe69} and can be formulated in both a strong sense (in a reasonable space-time we cannot have a naked singularity), or a weak sense (even if such singularities occur they are safely hidden behind an event horizon and cannot communicate with outside observers). Since Penrose' s initial proposal, there have been numerous attempts to try to prove the conjecture in a general sense (see \cite {Jo93} and references therein) but, unfortunately, none have been successful so far.\\
\indent
Since, due to the complexity of the full Einstein equations, the general problem appears (for the time being), to be intractable, metrics with specific symmetries must be used to construct concrete models of gravitational collapse whose properties, such as the existence of horizons, can then be determined. One such example is the two-dimensional reduction of general relativity obtained by imposing spherical symmetry. Even with this scenario, however, very few inhomogeneous, exact, non-static solutions have been found. One well known example is the Vaidya metric \cite{Va51}, which describes the gravitational field associated with the eikonal approximation of an isotropic flow of unpolarized radiation, or, in other words, a null fluid. It is asymptotically flat and is employed in modeling the external field of radiating stars and evaporating black holes. A second example is the Tolman-Bondi metric \cite{To34}, which gives the gravitational field associated with dust matter and is frequently applied either in cosmological models or in describing the collapse of a star to form a black hole. Tolman-Bondi spacetimes include the Schwarzschild solution, the Friedman-Robertson-Walker metric and the Oppenheimer-Snyder collapse, as well as inhomogeneous expansions and collapses, which may also lead to naked singularities \cite{DwJo91}.\\
\indent
At first sight these two metrics are completely different. Do the naked singularities that form in the collapse of null radiation and in the collapse of dust bear any relation to each other? Are there any features common to both solutions? And, if so, what are the implications for cosmic censorship? As shown by Lemos \cite{Le92}, the naked singularities which appear in the Vaidya and Tolman-Bondi space-times are of the same nature. Various important features such as the degree of inhomogeneity  necessary to produce a naked singularity, the Cauchy horizon equation, the apparent horizon equation, the strength of the singularity and the stability of the space-time have a mutual correspondence in both metrics. For cosmic censorship, this result
implies that, if the shell-focusing singularities arising from the collapse of a null fluid are not artifacts of some (eikonal) approximation, then the shell-focusing singularities arising from the collapse of dust are, likewise, not artifacts, and vice versa.
Conversely, if the naked singularities are artifacts of an eikonal approximation in one metric, they must also be artifacts in the other.\\
\indent
Thus the Vaidya solution belongs to the Tolman-Bondi family and the most unbound case yields the Vaydia metric, as originally discovered in \cite{Va51}. It is therefore reasonable to expect that major features which arise in one may also appear in the other. One example of such a result is the fact that the strength of singularity in the Vaidya metric depends on the direction from which the geodesics enter \cite{Le92}.\\
\indent
Null fluids, are, in principle, easier to treat than matter fields. A null fluid is the eikonal approximation of a massless scalar field. Thus if one shows that the naked singularities arising in the Vaidya metric can be derived from more fundamental (massless) fields, then the naked singularities which form in the Tolman-Bondi collapse may also be derived from more fundamental (massive) fields. The same types of relations and conclusions hold for charged radiation and charged pressureless matter (i.e. dust). The structure and properties of singularities arising from gravitational collapse in Vaidya space-times have been analyzed, from different points of view, in \cite{BoVa70,LaZa91,SuIs80,PoIs90,Or98}.\\
\indent
Within the framework of various physical models, spherically-symmetric gravitational collapse has been analyzed in many papers. The role of initial density and velocity distributions towards determining the final outcome of spherical dust collapse and the causal structure of the singularity has been examined in terms of the evolution of apparent horizons in \cite{SiJo96,JhJoSi96}. This collapse is described by the Tolman-Bondi metric with two free functions and can end in either in the formation of a black hole or a naked singularity. The occurrence and nature of naked singularities in the Szekeres space-times, representing irrotational dust, were investigated in \cite{JoKr96}. The Szekeres space-times have no Killing vectors and are the generalizations of the Tolman-Bondi solutions. Naked singularities that satisfy both the limiting focusing condition and the strong limiting focusing condition also exist. The role of the initial state of a collapsing dust cloud in determining its final fate has been considered in \cite{DwJo97}. For an arbitrary matter distribution at $t=0$, there is always the freedom to choose the rest of the initial data, namely the initial velocities of the collapsing spherical shells, so that, depending on this choice, the collapse
could result in either a black hole or a naked singularity. Thus, given the initial density profile, to achieve the desired end state of the gravitational collapse one has to give a suitable initial velocity distribution to the cloud. The expression for the expansion of outgoing null geodesics in spherical dust collapse was derived in \cite{Si97} and the limiting values of the expansion in the approach to singularity formation were computed. Using these results one can show that the horizon-shielded, as well as the naked singularity solutions arising in spherical dust collapse, are stable with respect to small perturbations in the equation of state.\\
\indent
The growth of the Weyl curvature is examined in two examples of naked singularity formation in spherical gravitational collapse (collapsing dust and the Vaidya space-time), in \cite{BaSi97}. The Weyl scalar diverges along outgoing, radial, null geodesics as they meet the naked singularity in the past. Although general relativity admits naked singularities arising from gravitational collapse, the second law of thermodynamics could forbid their occurrence in nature. A simple model for the corona of a neutrino-radiating star showing critical behavior is presented in \cite{Go98} and the conditions for the existence, or absence, of a bounce (i.e. explosion) are discussed. The charged Vaidya metric was extended to cover the whole space-time in \cite{PaWi98} and the Penrose diagram for the formation and evaporation of a charged black hole was therefore obtained. The covariant equations characterizing the strength of a singularity in spherically-symmetric space-times and a slight modification to the definition
of singularity strength were derived in \cite{No99}. The idea of probing naked space-time singularities with waves rather than with particles was also proposed in \cite{IsHo99}. For some space-times, the classical singularity becomes regular if probed
with waves, while stronger classical singularities remain singular.\\
\indent
In order to obtain the correct energy-momentum tensor for the collapse of a null fluid in a spherically-symmetric geometry, an inverted approach was proposed by Husain \cite{Hu96}. In this method, the general form of the stress-energy tensor is determined from the general form of the metric. The equation of state and an appropriate set of energy conditions, for example the dominant energy conditions (DEC), or weak energy conditions (WEC), are then imposed on its eigenvalues. This leads to a set of partial differential equations for the metric function, which determines the mass within a given radius, at any given time. The precise form of the stress-energy tensor is then displayed. Using this approach, two classes of solutions, describing the collapse of a null fluid satisfying the barotropic and polytropic equations of state, were obtained in  \cite{Hu96}. Within a similar framework, a large class of solutions, including those for the collapse of type II fluids \cite {LaLi75,ShTe83} and most of the known solutions of the Einstein field equations, was derived, in four dimensions, by Wang and Wu \cite{WaWu98}, and in $N\geq 4$ dimensions by Villas da Rocha \cite{Ro02}. The radiating Vaidya metric has also been extended to include both a radiation field and a string fluid by Glass and Krisch \cite{GlKr98,GlKr99} and by Govinder and Govender \cite{GoGo03}. \\
\indent
Further work on inhomogeneous, non-static, spherically-symmetric solutions in general relativity is presented in \cite{Goetal03,DaGh04,NaGoGo06}, including one class of solutions extended to include shear \cite{NaGoGo06}, and recent results regarding the formation of singularities, both naked and clothed, are contained in \cite{PaSaGh99,Lu13,Jo13,KoMaBa13}. Interestingly, the radiation spectra emitted from dust clouds collapsing to form both black holes and naked singularities are studied in \cite{KoMaBa13} and found to be in qualitative agreement, suggesting it may be difficult, even in principle, for observational tests to distinguish between the two, at least within the context of a four-dimensional Einstein universe. Similar classes of solutions, generalized to higher-dimensional space-times, are given in \cite{IyVi89,PaDa99,GhDa01}.\\
\indent
When nuclear matter is compressed to a sufficiently high density, a phase transition is thought to occur which converts neutron matter into free, three-flavor, (strange) quark matter, due to the fact that the latter is expected to be more stable than the former. The collapse of the quark fluid, described by the bag model equation of state, $p=\left(\rho - 4B\right)/3$, with $B =$ const. $\approx 10^{14}$ gcm${}^{-3}$, was studied in general relativity by Harko and Cheng \cite{HaCh00} and
the conditions for the formation of a naked singularity were obtained. This solution was later generalized to arbitrary space-time dimensions and to a more general linear equation of state by Ghosh and Dadhich \cite {GhDh02,GhDh03}. The nucleation of quark matter in neutron star cores, again in the context of general relativity, was considered in \cite{HaChTa04} and observational consequences of quark star remnants (within the same paradigm) are discussed in \cite{Chetal09}. Quark-Hadron phase transitions in brane world models were also studied in \cite{DeRiHa09}.\\
\indent
In 1965 Hagedorn \cite{Ha65} postulated that, for large masses $m$, the spectrum of hadrons, $\rho \left( m\right)$, grows exponentially according to $\rho \left( m\right) \sim \exp \left(m/T_{H}\right) $, where $T_{H}$, known as the Hagedorn temperature, is a scale parameter. The hypothesis was based on the observation that, at some point, a further increase of energy in proton-proton and proton-antiproton collisions no longer raises the temperature of the resulting fireball but, instead, results in increased particle production. Thus there is a maximum temperature $T_{H}$ that a hadronic system can achieve. This statistical model of hadrons has been used, phenomenologically, to describe matter at densities exceeding the nuclear density \cite{Br00,Br04,Ha70,Rh71}. A Hagedorn-type  phase transition also occurs naturally in theories containing fundamental strings, since these have a large number of internal degrees of freedom \cite{Gi89}. As a result of the existence of many oscillatory modes, the density of states grows exponentially with the energy of a single string. Thermodynamical quantities, such as the entropy, diverge at the Hagedorn temperature. If one considers an ensemble of weekly interacting strings at a finite temperature, the behavior of the density of states is thought to lead either to a limiting temperature, or to a phase transition in which the string configuration changes to one which is dominated by a single long string \cite{Gr01}. The high density Hagedorn phase of matter has also been extensively used in cosmology to describe the very early phases of the evolution of the Universe \cite{Ma98,Be02,Ma03,Ba03} and the spherically-symmetric collapse of a Hagedorn fluid in the Vaidya geometry has been studied in \cite{Ha03}.\\
\indent
The aim of the present work is to study the spherically-symmetric collapse of various types of cosmological and astrophysical fluid in the brane world scenario. In order to simplify the mathematical formalism, we adopt the assumption that the high density fluid moves along the null geodesics of a Vaidya type space-time. The Vaidya geometry, which also permits the incorporation of radiative effects, offers a more realistic background than static geometries, where all back reaction is ignored. \\
\indent
The structure of the paper is then as follows. In Sect. \ref{field} we review the basic mathematical formalism of brane world models and determine the conditions under which non-local bulk effects may be ignored, leading to isotropic, quadratic, corrections to the effective energy density and pressure on the brane. The effective values are found to be functions of only the physical energy density and pressure of the brane world matter but also depend on the value of the (constant) brane tension.  In Sect. \ref{Section3} we consider the high density limit of these functions, before substituting an appropriate equation of state to obtain the (approximate) brane world field equations. Exact solutions to the approximate field equations are then found, in terms of two arbitrary integration functions, and an appropriate set of energy conditions is applied, constraining their general forms. The behavior of null geodesics is then investigated in order to determine the equation which governs the formation of apparent horizons  for each fluid type. This, in turn, allows us to specify the conditions necessary for the formation of naked singularities in each collapse scenario. Sections \ref{Section3.1}-\ref{Section3.4} deal with null fluids satisfying the barotropic, polytropic, strange quark  ``bag" model and Hagedorn equations of state, respectively. As an astrophysical application of this work, in Section~\ref{Section3.5} we consider the possibility that energy emission from a Hagedorn fluid, collapsing to form a naked singularity, may be a source of GRBs in the brane world and compare these results to those obtained from a similar analysis in four-dimensional general relativity  \cite{Ha03}. Section~\ref{Section4} contains a summary of our results and a brief discussion of their overall relation to their general-relativistic counterparts.

\section{The field equations for static, spherically-symmetric vacuum branes}

\label{field}

In this Section we briefly review the basic mathematical formalism of brane world models, present the field equations for a static, spherically-symmetric vacuum brane, and discuss some of their consequences. We begin by considering a single, four-dimensional (4D) brane, on which matter is confined, embedded within a five-dimensional (5D) bulk. Thus, the brane world $({}^{(4)}M,g_{\mu \nu })$ is localized on a hyper-surface $(B( X^{A}) =0$,  with coordinates $X^{A},  \ A \in \left\{0,1,2,3,4\right\})$ in the bulk space-time $({}^{(5)}M,g_{AB})$. The induced 4D coordinates on the brane are labelled $x^{\mu }, \ \mu \in \left\{0,1,2,3,4\right\}$.

The action of the system is given by~\cite{Shiromizu}
\begin{equation}
S=S_{bulk}+S_{brane},  \label{bulk}
\end{equation}
where
\begin{equation}
S_{bulk}=\int_{{}^{(5)}M}\sqrt{-{}^{(5)}g}\left[ \frac{1}{2k_{5}^{2}}{}%
^{(5)}R+{}^{(5)}L_{m}+\Lambda _{5}\right] d^{5}X,
\end{equation}
and
\begin{equation}
S_{brane}=\int_{{}^{(4)}M}\sqrt{-{}^{(5)}g}\left[
\frac{1}{k_{5}^{2}}K^{\pm }+L_{brane}\left( g_{\alpha \beta },\psi
\right) +\lambda _{b}\right] d^{5}X,
\end{equation}
where $k_{5}^{2}=8\pi G_{5}$ is the 5D gravitational constant, ${}^{(5)}R$ and ${}^{(5)}L_{m}$ are the 5D scalar curvature and the matter Lagrangian in the bulk, $L_{brane}\left( g_{\alpha \beta },\psi \right) $ is the 4D Lagrangian, which is given by a generic functional of the brane metric $g_{\alpha \beta }$ and the matter fields $\psi $, $K^{\pm }$ is the trace of the extrinsic curvature on either side of the brane, and $\Lambda _{5}$ and $\lambda _{b}$ (the constant brane tension) are the negative vacuum energy densities in the bulk and on the brane, respectively.

The Einstein field equations in the bulk are given by~\cite{Shiromizu}
\begin{equation}
{}^{(5)}G_{IJ}=k_{5}^{2} {}^{(5)}T_{IJ},
\ee
where
\bea
&&{}^{(5)}T_{IJ}\equiv  \frac{-2}{\sqrt{-{}^{(5)}g}}\frac{\delta ({}^{(5)}L_{m}\sqrt{-{}^{(5)}g})}{\delta
{}^{(5)}g^{IJ}}=\nonumber\\
&&-\Lambda _{5}
{}^{(5)}g_{IJ}+\delta(B)\left[-\lambda_{b}
{}^{(5)}g_{IJ}+T_{IJ}\right] ,
\eea
is the energy-momentum tensor of the bulk matter fields, while $T_{\mu\nu }$ is the energy-momentum tensor localized on the brane, which is defined by
\begin{equation}
T_{\mu \nu }\equiv \frac{-2}{\sqrt{-g}}\frac{\delta (L_{brane}\sqrt{-g})}{\delta g^{\mu \nu
}}.
\end{equation}

The delta function $\delta \left(B\right) $ denotes the localization of brane contribution. In the 5D space-time the position of the brane represents the fixed point of a $Z_{2}$ symmetry and the ``on brane" values of basic fields are obtained by projections from the bulk onto the 4D hyper-surface it occupies. In particular, the induced 4D metric is given by $g_{IJ}={}^{(5)}g_{IJ}-n_{I}n_{J}$, where $n_{I}$ is the space-like unit vector field normal to the brane hyper-surface, ${}^{(4)}M$. In the following we assume that the matter content of the bulk, outside the brane hyper-surface is zero, ${}^{(5)}L_{m}=0$, for the sake of simplicity. Since the entire matter content of the theory is confined to the 4D brane world, in this scenario, only gravity can probe the extra dimensions.

Assuming a metric of the form $ds^{2}=(n_{I}n_{J}+g_{IJ})dx^{I}dx^{J}$, where $n_{I}$ denotes the unit normal to the set of $\chi =\mathrm{const.}$ hyper-surfaces, $n_{I}dx^{I}=d\chi$, and $g_{IJ}$ is the induced metric on these hyper-surfaces, the effective 4D gravitational equation on the brane takes the form~\cite{Shiromizu}:
\begin{equation}
G_{\mu \nu }=-\Lambda g_{\mu \nu }+k_{4}^{2}T_{\mu \nu
}+k_{5}^{4}S_{\mu \nu }-E_{\mu \nu },  \label{Ein}
\end{equation}
where $S_{\mu \nu }$ is the local quadratic energy-momentum correction
\begin{equation}
S_{\mu \nu }=\frac{1}{12}TT_{\mu \nu }-\frac{1}{4}T_{\mu
}{}^{\alpha }T_{\nu \alpha }+\frac{1}{24}g_{\mu \nu }\left(
3T^{\alpha \beta }T_{\alpha \beta }-T^{2}\right) ,
\end{equation}
and $E_{\mu \nu }$ is the non-local effect from the free bulk gravitational field, given by the transmitted projection of the bulk Weyl tensor ($C_{IAJB}$), $E_{IJ}=C_{IAJB}n^{A}n^{B}$, which obeys the property $E_{IJ}\rightarrow E_{\mu \nu }\delta _{I}^{\mu }\delta _{J}^{\nu }\quad $as$\quad \chi \rightarrow 0$. Here, $k_4$ denotes the 4D coupling constant, which is related to the usual, 4D, gravitational constant via $k_{4}^{2}=8\pi G$.

The effective 4D and fundamental 5D cosmological constants, $\Lambda$ and $\Lambda_{5}$, and the 4D and 5D coupling constants, $k_{4}$ and $k_{4}$, are related to each other, and to the intrinsic brane tension, $\lambda_{b}$, via $\Lambda =k_{5}^{2}(\Lambda_{5}+k_{5}^{2}\lambda _{b}^{2}/6)/2$ and $k_{4}^{2}=k_{5}^{4}\lambda _{b}/6$, respectively. Therefore, in the limit $\lambda _{b}^{-1}\rightarrow 0$, we recover the action and equations of motion of standard general relativity on the brane \cite{Shiromizu}.

The brane tension $\lambda _b$ is an important parameter in determining the properties of brane world models. Its numerical value is constrained by the validity of Newtonian gravity, in four dimensions, on length scales smaller than 0.1 mm \cite{Hoyle}. The minimum bound on $\lambda _b$ from such experiments is given by $\lambda _b\geq \left (100\;{\rm GeV}\right)^4$ \cite{Hamed,Chung}. If certain stringent conditions on the brane tension were satisfied, post-inflationary brane cosmology would always deviate from the standard Big Bang cosmology. For example, the necessity of nucleosynthesis on the brane leads to a constraint on the brane tension of $\lambda _b > \left(1\;{\rm MeV}\right)^4$ \cite{Mazumdar}.  Furthermore, if we do not want to plague the inflaton potential with non-renormalizable quantum corrections, inflation is required to begin at a scale below the four-dimensional Planck mass, thus constraining the brane tension to $10^{64}\;({\rm GeV})^4<\lambda _b<1\;({\rm MeV})^4$ \cite{Mazumdar}. In brane world models the reheating temperature $T_{reh}$ satisfies, to a very good approximation, the relation $T_{reh}\approx \lambda _b^{1/4}$ \cite{Allahverdi, Harko} and this imposes the constraint $0.891 \times 10^{25} \;({\rm GeV})^4 \geq \lambda _b \geq 14.26 \times 10^{25}\; ({\rm GeV})^4$ \cite{Harko}.

The five-dimensional gravitational coupling constant $k_5$ and the five-dimensional cosmological constant $^{(5)}\Lambda$ are constrained by the present value of the gravitational constant $G$ via $\left(k_5^4/6\right)\sqrt{\lambda _b^2-6 ^{(5)}\Lambda /k_5^2}\approx k_5^3\sqrt{-^{(5)}\Lambda /6}=8\pi G\approx 1.68\times 10^{-55} \;{\rm eV}^{-2}$, while the five-dimensional cosmological constant is also constrained by $\left(k_5^2/2\right)\;^{(5)}\Lambda \approx -6/\left(0.1\;{\rm mm}\right)^2\approx -2.3\times 10^{-5} \;{\rm eV}^2$ \cite{MaKo10}. From these two conditions, we obtain $k^4_5 \approx  3.6 \times 10^{-105} {\rm eV}^{-6}$ and $^{(5)}\Lambda \approx -7.7 \times 10^{46}\; {\rm eV}^5$. In addition, the value of the brane tension $\lambda _b$  can be estimated from the value of the present day dark energy density, $\rho _{dark}\approx 10^{-12}\;{\rm eV}^4$ \cite{Planck} via $k_5^4\lambda _b/12=8\pi G \rho _{dark}\approx 1.6\times 10^{-67}\;{\rm eV}^2$, giving $\lambda _b\approx 1.6\times 10^{19}\;{\rm eV}^4$ \cite{Harko1}.

The Einstein equation in the bulk, together with the Codazzi equation, also imply the conservation of the energy-momentum tensor for the matter on the brane, $D_{\nu }T_{\mu }{}^{\nu }=0$, where $D_{\nu }$ denotes the brane
covariant derivative. Moreover, from the contracted Bianchi identities on the brane, it follows that the projected Weyl tensor obeys the constraint $D_{\nu }E_{\mu }{}^{\nu }=k_{5}^{4}D_{\nu}S_{\mu }{}^{\nu }$.

The symmetry properties of $E_{\mu \nu }$ imply that, in general, we can decompose it irreducibly with respect to a chosen $4$-velocity field, $u^{\mu}$, according to~\cite{Mart04}
\begin{equation}
E_{\mu \nu }=-\kappa^{4}\left[ U\left( u_{\mu }u_{\nu}
+\frac{1}{3}h_{\mu \nu }\right) +P_{\mu \nu }+2Q_{(\mu
}u_{\nu)}\right] ,  \label{WT}
\end{equation}
where $\kappa=k_{5}/k_{4}$ and $h_{\mu \nu }=g_{\mu \nu }+u_{\mu }u_{\nu}$ projects orthogonal to $u^{\mu }$. Here, $U=-\kappa^{-4}E_{\mu \nu }u^{\mu }u^{\nu }$ is a scalar  ``dark radiation'' term, $Q_{\mu}=\kappa^{-4}h_{\mu }^{\alpha }E_{<\alpha >\beta }u^{\beta }$ is a spatial vector and $P_{\mu \nu }=-\kappa^{-4}\left[ h_{(\mu }\text{}^{\alpha }h_{\nu )}\text{ }^{\beta }-\frac{1}{3}h_{\mu \nu
}h^{\alpha \beta }\right] E_{\alpha \beta }$ is a spatial, symmetric, and trace-free tensor.

For the vacuum state, we have $\rho =p=0$, so that $T_{\mu \nu}\equiv 0$ and, consequently, $S_{\mu \nu }\equiv 0$. In this case the field equation describing a static brane takes the form
\begin{equation}
R_{\mu \nu }=-E_{\mu \nu }+\Lambda g_{\mu \nu },
\end{equation}
with the trace, $R$, of the Ricci tensor, $R_{\mu \nu }$, satisfying the condition $R=R_{\mu }{}^{\mu }=4\Lambda $.

In the vacuum case, the constraint on $E_{\mu \nu }$ reduces to $D_{\nu}E_{\mu}{}^{\nu }=0$. Since, in an inertial frame at any point on the brane we have that $u^{\mu }=\delta _{0}^{\mu }$ and $h_{\mu \nu}=\mathrm{diag}(0,1,1,1)$, for a static vacuum, $Q_{\mu }=0$, the constraint takes the form~ \cite{GeMa01}
\begin{equation}
\frac{1}{3}D_{\mu }U+\frac{4}{3}UA_{\mu }+D^{\nu }P_{\mu \nu
}+A^{\nu }P_{\mu \nu }=0,
\end{equation}
where $A_{\mu }=u^{\nu }D_{\nu }u_{\mu }$ is the 4-acceleration.
In the static, spherically-symmetric, case we may choose $A_{\mu }=A(r)r_{\mu }$ and $P_{\mu \nu }=P(r)\left( r_{\mu }r_{\nu }-\frac{1}{3}h_{\mu \nu}\right) $, where $A(r)$ and $P(r)$ (the ``dark pressure'') are some scalar functions of the radial distance~$r$, and~$r_{\mu }$ is a unit radial vector~\cite{Da00}.

The general form of the brane energy-momentum tensor for any matter fields (scalar fields, perfect fluids, kinetic gases, dissipative fluids etc.), including a combination of different fields, can be written covariantly as
\begin{equation}
T_{\mu \nu }=\rho u_{\mu }u_{\nu }+ph_{\mu \nu }+\pi _{\mu \nu
}+q_{\mu }u_{\nu }+q_{\nu }u_{\mu },
\end{equation}
where $\rho $ and $p$ are the energy density and the isotropic pressure, respectively. The energy flux obeys $q_{\mu}=q_{\left\langle \mu \right\rangle }$, and the anisotropic stress obeys $\pi _{\mu \nu }=\pi _{\left\langle \mu \nu \right\rangle }$, where the angled brackets denote the projected, symmetric and trace-free part, so that $V_{\left\langle \mu \right\rangle }=h_{\mu }^{\nu }V_{\nu }$ and $W_{\left\langle \mu \nu \right\rangle }=\left[ h_{(\mu }^{\alpha}h_{\nu )}^{\beta }-(1/3)h^{\alpha \beta }h_{\mu \nu }\right] W_{\alpha \beta }$, respectively, with round brackets denoting symmetrization. Therefore, in an inertial frame at any point on the brane, we have that $q_{\mu }=\left(0,q_{i}\right) $ and $\pi _{\mu 0}=0$.

The local and non-local bulk corrections may be combined into an effective total energy-density, pressure, anisotropic stress and energy-flux, so that the modified field equations of the brane world models can be written in the standard form
\begin{equation} \label{sf}
G_{\mu \nu }=-\Lambda g_{\mu \nu }+k_4^{2}T_{\mu \nu }^{eff}\text{,}
\end{equation}
where the components of $k_4^{2}T_{\mu \nu }^{eff}$ are
\begin{equation}
\rho ^{eff}=\rho +\frac{1}{4\lambda_b }\left( 2\rho ^{2}-3\pi _{\mu
\nu }\pi ^{\mu \nu }\right) +\frac{6}{\kappa^{4}\lambda_b }U,
\end{equation}
\begin{equation}
p^{eff}=p+\frac{1}{4\lambda_b }\left( 2\rho ^{2}+4\rho p + \pi _{\mu
\nu }\pi ^{\mu \nu }-4q_{\mu }q^{\mu }\right)
+\frac{2}{\kappa^{4}\lambda_b }U,
\end{equation}
\bea
\pi _{\mu \nu }^{eff}&=&\pi _{\mu \nu }+\frac{1}{2\lambda_b }\left[
-\left( \rho +3p\right) \pi _{\mu \nu }+\pi _{\alpha <\mu }\pi
_{\nu >}^{\alpha }+q_{<\mu }q_{\nu >}\right]+\nonumber\\
&&\frac{6}{\kappa^{4}\lambda_b }P_{\mu \nu },
\eea
\begin{equation}
q_{\mu }^{eff}=q_{\mu }+\frac{1}{4\lambda_b }\left( 4\rho q_{\mu
}-\pi _{\mu \nu }q^{\nu }\right) +\frac{6}{\kappa^{4}\lambda_b }Q_{\mu }.
\end{equation}

For a perfect fluid, we have that $q_{\mu }=0$, $\pi _{\mu \nu }=0$, and, in this case
\begin{equation}
q_{\mu }^{eff}=\frac{6}{k^{4}\lambda_b }Q_{\mu },\pi _{\mu \nu }^{eff}=\frac{6}{\kappa^{4}\lambda_b }P_{\mu \nu }.
\end{equation}

Finally, assuming that the bulk curvature is small, we may also neglect terms in $P_{\mu\nu}$, $Q_{\mu}$ and $U$, so that
\begin{eqnarray} \label{eq:rho_eff}
\rho^{eff} \approx \rho + \frac{1}{2\lambda_b} \rho^2,
\end{eqnarray}
\begin{eqnarray} \label{eq:p_eff}
p^{eff} \approx \rho + \frac{1}{2\lambda_b} \rho^2 + \frac{1}{\lambda_b} \rho p,
\end{eqnarray}
\begin{equation}  \label{eq:q_eff}
q_{\mu }^{eff} \approx \pi _{\mu \nu }^{eff} \approx 0.
\end{equation}
Since the anisotropic stress is zero and the effective energy density and pressure on the brane, $\rho^{eff}$ and $p^{eff}$, are then functions only of the physical energy density and pressure of the brane world matter, $\rho$ and $p$, and of the constant brane tension, $\lambda_b$, the resulting field equations simplify considerably. In the high density limit, where we may also neglect linear terms in $\rho$, their form is simplified even further. In the next Section, we will use the formalism and approximations outlined above to obtain exact solutions to the (approximate) field equations, obtained by substituting Eqs. (\ref{eq:rho_eff})-(\ref{eq:q_eff}) into Eq. (\ref{sf}), for four types of null fluid of particular astrophysical or cosmological interest.

\section{Collapsing null fluids in the brane world models} \label{Section3}

In ingoing Bondi coordinates $(u,r,\theta ,\varphi )$ and using the advanced Eddington time coordinate, $u=t+r$ (with the radial coordinate $r\geq 0$ decreasing towards the future), the line element describing the radial collapse of a coherent stream of matter can be represented in the form \cite {Hu96}, \cite{HaCh00}
\begin{equation}
ds^{2}=-\left[ 1-\frac{2m\left( u,r\right) }{r}\right] du^{2}+2dudr+r^{2}%
\left( d\theta ^{2}+\sin ^{2}\theta d\varphi ^{2}\right) .
\label{222}
\end{equation}
where $m\left(u,r\right) $ is the mass function, which gives the gravitational mass within a given radius $r$ for any value of $u$. In the following we use the natural system of units with $8\pi G=c=1$.

The matter energy-momentum tensor on the brane can be written in the form \cite{Hu96,WaWu98}
\begin{equation}
T_{\mu \nu }^{eff}=T_{\mu \nu }^{(n)}+T_{\mu \nu }^{(m)},
\label{32}
\end{equation}
where
\begin{equation}
T_{\mu \nu }^{(n)}=\mu ^{eff}\left( u,r\right) l_{\mu }l_{\nu },
\end{equation}
is the component of the matter field that moves along the null hypersurfaces $u=const.$, and
\begin{equation}
T_{\mu \nu }^{(m)}=\left( \rho ^{eff}+p^{eff}\right) \left( l_{\mu
}n_{\nu }+l_{\nu }n_{\mu }\right) +p^{eff}g_{\mu \nu },
\label{52}
\end{equation}
represents the energy-momentum tensor of the collapsing matter. Here, $l_{\mu }$ and $n_{\mu }$ are two null vectors, given by $l_{\mu}=\delta _{\left( \mu \right) }^{\left(0\right) }$ and $n_{\mu }=\frac{1}{2}\left[ 1-\frac{2m\left( u,r\right) }{r}\right] \delta _{\left( \mu \right) }^{\left( 0\right) }-\delta_{\left( \mu \right) }^{\left( 1\right) }$, so that $l_{\alpha}l^{\alpha }=n_{\alpha }n^{\alpha }=0$ and $l_{\alpha }n^{\alpha}=-1$. The energy density and pressure in Eq. (\ref{52}) are obtained by diagonalizing
the energy-momentum tensor obtained from the metric \cite{Hu96}.

For the effective energy-momentum tensor given in Eq. (\ref{32}) the gravitational field equations take the general form \cite{HaCh00}
\begin{equation}
\frac{2}{r^{2}}\frac{\partial m\left( u,r\right) }{\partial u}=
\mu ^{eff}\left( u,r\right) ,  \label{rad}
\end{equation}
\begin{equation}
\frac{2}{r^{2}}\frac{\partial m(u,r)}{\partial r}=\rho
^{eff}\left( u,r\right) ,  \label{dens}
\end{equation}
\begin{equation}
-\frac{1}{r}\frac{\partial ^{2}m\left( u,r\right) }{\partial
r^{2}}=p^{eff}\left( u,r\right).  \label{pres}
\end{equation}

Taking the general assumptions made in deriving Eqs. (\ref{eq:rho_eff})-(\ref{eq:p_eff}), namely that the collapsing matter is a perfect fluid and that the bulk curvature is small, the equations of motion then become
\begin{equation}
\frac{2}{r^{2}}\frac{\partial m(u,r)}{\partial r} \approx \rho + \alpha \rho^2,  \label{radf}
\end{equation}
\begin{equation}
-\frac{1}{r}\frac{\partial ^{2}m\left( u,r\right) }{\partial r^{2}} \approx \rho + \alpha \rho^2 +2\alpha \rho p. \label{presf}
\end{equation}
where
\begin{equation} \label{alpha}
\alpha = \frac{1}{2\lambda_b} = \frac{1}{12}\frac{k_5^4}{k_4^2}.
\end{equation}
The next step is to impose an equation of state relating $p$ and $\rho$. The field equations may then be solved and the form of $m(u,r)$ determined up to arbitrary functions of $u$. This gives us a general model for the spherically-symmetric collapse of a null fluid governed by our chosen equation of state in which specific solutions can be chosen by specifying the forms of the remaining functions. In practice however, we may first simplify the above field equations even further by considering the high density limit $\rho \rightarrow \infty$ in which $\alpha \rho^2 >> \rho$. This assumption is valid in most astrophysical situations and allows us to neglect linear terms in $\rho$. Substituting the expression for $m(u,r)$ back into the metric then allows us to investigate the properties of outgoing null geodesics near $r=0$ and to calculate the positions of any horizons which may be present. In particular we would like to know if brane world corrections to the standard field equations favour the formulation of naked singularities or black holes, in contrast to the corresponding cases in four-dimensional general relativity, and if any violations of the cosmic censorship conjecture may occur. Naively however, we would expect gravity to be stronger on smaller scales in the brane world model, therefore favouring the formation of black holes as opposed to naked singularities. We must also impose an appropriate set of energy conditions on the components of the energy momentum tensor, which imply constraints on the form of $m(u,r)$. The stress-energy tensor Eq. (\ref{52}) satisfies the dominant energy
condition (DEC), if the following three conditions are met:
\begin{equation} \label{DEC}
p^{eff}\geq 0, \ \  \rho ^{eff}\geq p, \ \ T_{\mu\nu}^{eff}w^{\mu}w^{\nu}\geq0,
\end{equation}
where $w^{a}$ is an arbitrary time-like (or null) four-vector. The first two of these conditions imply that $\frac{\partial m}{\partial r}\geq 0$ and $\frac{\partial ^{2}m}{\partial r^{2}}\leq 0$. The former just says that the mass function either increases with $r$ or is a constant, which is a natural physical requirement. Substituting $w^{\mu} = n^{\mu}$, we see that the final condition implies $\mu^{eff}(u,r)\geq 0$, which in turn requires $\frac{\partial m}{\partial u} \geq 0$. A less stringent set of constraints, known as the weak energy conditions (WEC), which represent the minimal physical requirements, are
\begin{equation} \label{WEC}
p^{eff}\geq 0, \ \ \rho ^{eff}\geq 0.
\end{equation}
To satisfy either the WEC or the DEC one must therefore impose an appropriate equation of state for the collapsing matter.

In the following four Sections we investigate the collapse of four types of null fluid in the high density limit. Sections~\ref{Section3.1} and \ref{Section3.2} deal with ``generic" null fluids governed by the one-parameter barotropic and two-parameter polytropic equations of state with parameters $0 \leq k \leq 1$ and $0 \leq k \leq 1$, $a>0$, ($a \neq 1$), respectively.
In Section~\ref{Section3.3} we investigate collapsing strange quark matter using the MIT bag model. As mentioned above, this case is of particular astrophysical interest as free strange matter is expected to be the most stable form of matter at the high temperatures present in the cores of massive neutron stars. Finally, in Section~\ref{Section3.4}, we investigate the collapse of a Hagedorn fluid. The Hagedorn equation of state may be used to model ordinary matter at energy densities much higher than the nuclear density $\rho_n \approx 2 \times 10^{14}$ gcm$^{-3}$. As also stated in the Introduction, in the Hagedorn scenario increasing the energy density only increases the temperature of the system up to a certain maximum temperature known as the Hagedorn temperature, $T_H \approx 150-190$ MeV. Increasing the energy density beyond this point leads to the production of new particles but not to an increase in the kinetic energy of the system due to the presence of a large number of baryonic resonances at high energies. The Hagedorn equation of state may therefore be appropriate for describing ordinary matter at high densities in the early universe and the dynamics of a collapsing Hagedorn fluid could have important cosmological implications.

In each of the four cases, since the gravitational  field equations are second order, each model depends on two arbitrary functions, which we label $f_i(u)$, and $g_i(u)$, where $i \in \{b,p,s,H\}$ denotes barotropic, polytropic, strange quark and Hagedorn, respectively. Since it is also instructive to compare the solutions obtained for the brane world with the corresponding solutions for each type of fluid in general relativity, we will do so in each section. In order to avoid confusion between quantities and expressions in general relativity and in the brane world, we will denote the former with the additional subscript GR. In cases where additional emphasis is desirable, we will also denote the latter with the subscript BW, though this will be dropped where such emphasis is unnecessary.

\section{Solution of the field equations for the barotropic equation of state} \label{Section3.1}

Given the field equations, Eqs. (\ref{rad})-(\ref{pres}), we now impose the one-parameter barotropic equation of state,
\begin{equation}\label{barotropic}
p = k \rho
\end{equation}
where $0 \leq k \leq 1$ is a dimensionless parameter. In the high density limit $\rho \rightarrow \infty$ we have that $\alpha \rho^2 >> \rho$ and neglecting linear terms in $\rho$ leads to the approximate, but easily solvable equations
\begin{equation}\label{dens_A}
\frac{2}{r^2}\frac{\partial m(u,r)}{\partial r}  = \alpha \rho^2
\end{equation}
\begin{equation}\label{pres_A}
-\frac{1}{r}\frac{\partial^2 m(u,r)}{\partial r^2}  = (2k + 1)\alpha \rho^2.
\end{equation}
Substituting for $\alpha \rho^2$ from Eq. (\ref{dens_A}) into Eq. (\ref{pres_A}) gives
\begin{eqnarray}
\int \frac{\partial^2 m}{\partial r^2}  \left(\frac{\partial m}{\partial r}\right)^{-1} dr = -2(2k + 1)\int \frac{dr}{r},
\end{eqnarray}
yielding
\begin{equation} \label{eq:m'A}
\frac{\partial m(u,r)}{\partial r}  = \frac{g_b(u)}{r^{4k + 2}},
\end{equation}
where $g_b(u)$ is an arbitrary function of $u$. Integrating once more we obtain an explicit expression for $m(u,r)$ in terms of two arbitrary functions, $f_b(u)$ and $g_b(u)$,
\begin{equation}\label{eq:m_r_v1}
m(u,r) = f_b(u) - \frac{g_b(u)}{(4k + 1)r^{4k + 1}}.
\end{equation}
It is instructive at this point to compare the solution in Eq. (\ref{eq:m'A}) to that obtained in the corresponding case for general relativity. For a perfect fluid governed by the barotropic equation of state, Hussain \cite{Hu96} obtained the solution
\begin{equation} \label{Hussain_b}
m_{(GR)}(u,r) =  \left \lbrace
\begin{array}{rl}
f_{b}(u) - \frac{g_{b}(u)}{(2k - 1)r^{2k - 1}}, & k \neq 1/2 \\
f_{b}(u)+ g_{b}(u)\ln(r), & k=1/2
\end{array}\right.
\end{equation}
(using the notation adopted here). Note that in the brane world scenario $4k + 1 \neq 0$, for all $k$ in the range $0 \leq k \leq 1$, so that we have no need to consider a critical case in which neither the dominant nor the weak energy conditions can be satisfied. By contrast this occurs in the general relativistic solution when $k = 1/2$. As we will see, in the brane world model, both sets of energy conditions can always be satisfied for all values of $k$ and we now relate the energy density $\rho(u,r)$ and pressure $p(u,r)$ to the arbitrary functions which result from the integration of the field equations by imposing either the DEC or the WEC. Using the first field equation, Eq. $(\ref{dens_A})$, and the solution given in Eq. (\ref{eq:m'A}), we see that
\begin{equation} \label{eq:rho_r_v1}
\rho(u,r) = \pm \sqrt{\frac{2g_b(u)}{\alpha r^{4k + 4}}}.
\end{equation}
The physical requirement that $\rho(u,r)$ is real and positive, $\rho(u,r) \geq 0$, which is common to both sets of energy conditions, then implies that $g_b(u) \geq 0$, and that we must take the positive square root in the equation above. The WEC are therefore automatically satisfied for the barotropic equation of state for all possible values of $k$, $0 \leq k \leq 1$, by imposing $g_b(u) \geq 0$, which ensures that $\rho(u,r)$ is real and positive, while the form of $f_b(u)$ is left unconstrained. This result mirrors that obtained in general relativity \cite{Hu96}, namely that $g_b(u) \geq 0$ is sufficient to ensure positive energy density and pressure, thus satisfying the WEC, although slightly different arguments have been used. The second part of the DEC, $\rho^{eff}(u,r) \geq p(u,r)$, requires $\rho(u,r) \geq k/\alpha$, or
\begin{equation}
g_b(u) \geq \frac{1}{2}k^2 r^{4k+4},
\end{equation}
so that, if $R(u)$ represents the time-dependent maximal radius of the collapsing barotropic fluid, we have
\begin{equation} \label{DEC_pt2}
g_b(u) \geq \frac{1}{2}k^2R(u)^{4k+4}.
\end{equation}
This condition has no analogue in general relativity, since the DEC there simply require, $\rho(u,r) \geq p(u,r)$, which is automatically satisfied for $\rho(u,r) \geq 0$, ($g_b(u,r) \geq 0$) for a fluid obeying the barotropic equation of state. However, by Eq. (\ref{DEC_pt2}) we see that the function $g_b(u)$ characterizes the time-dependent radius of the collapsing sphere. Imposing the third component of the DEC results in the condition $\mu^{eff}(u,r) = (2/r^2)\partial m(u,r)/\partial r \geq 0$, or equivalently
\begin{equation} \label{DECA}
\frac{\partial f_b(u)}{\partial u} \geq \frac{1}{(4k + 1) r^{4k+1}} \frac{\partial g_b(u)}{\partial u}.
\end{equation}

In principle this constraint still allows for cases in which the derivatives of both functions are either positive or negative (or zero); $\partial f(u)/\partial u \geq 0$, $\partial g(u)/\partial u \geq 0$ or $\partial f(u)/\partial u \leq 0$, $\partial g(u)/\partial u \leq 0$, so long as the inequality is still obeyed. However, by far the simplest way to satisfy Eq. (\ref{DECA}) is by setting $\partial f(u)/\partial u \geq 0$ and $\partial g(u)/\partial u \leq 0$, for all $u$. The general class of space-times corresponding to the spherically symmetric-collapse of a null fluid, governed by the barotropic equation of state, Eq. (\ref{barotropic}), in the brane world are therefore described by metrics of the form
\bea \label{baro_metric}
ds^2 &=& -\left[1 - \frac{2f_b(u)}{r} + \frac{2g_b(u)}{(4k+1)r^{4k+2}} \right]du^2 + \nonumber\\
&&2dudr + r^2(d\theta^2 + \sin^2\theta d\phi^2),
\eea
where the functions $f_b(u)$ and $g_b(u)$ determine both the initial (generally inhomogeneous) distribution of matter and the time-dependent dynamics of the collapse. This, in turn, determines the key properties of the space-time, including the strength of the singularity, the positions of any horizons and the asymptotic geometry.

As in the case of spherically-symmetric collapse in general relativity we see that setting $\rho(u,r) = p(u,r) = 0$ implies $g_b(u) = 0$ which results the Vaidya metric. This metric is recovered in the brane world model when $\rho(u,r) = 0$, as expected, because the $\rho^2(u,r)$ terms which contribute the brane world corrections to the field equations vanish, along with the neglected linear terms. It is interesting to note that we may also reproduce the charged Vaidya metric by setting $k = 0$. This corresponds physically to the collapse of a fluid with positive energy density and zero pressure (i.e. a null dust) and contrasts with the general relativistic solution in which $k=1$ is required. That is, the collapse of a null dust with $\rho(u,r) \geq 0$, $p(u,r) = 0$ in the brane world scenario yields the same space-time, described by the charged Vaidya metric, as the collapse of a null fluid with $\rho(u,r) = p(u,r) \geq 0$ in general relativity. In particular the Reisner-Nordstrom metric may be recovered by the choosing $f_b(u) = M$, $2g_b(u) = Q^2$ and $k=0$, as opposed to $k=1$.

In general, however, at time $u$, the singularity at the center of the collapsing region will have multiple horizons given by the positive, real solutions of
\begin{eqnarray}
r^{C_{BW}+1} - f_b(u)r^{C_{BW}} + \frac{g_b(u)}{C_{BW}} = 0,
\end{eqnarray}
where $C_{BW}=4k+1$, whereas the analogous equation in general relativity is
\begin{eqnarray}
r^{C_{GR}+1} - f_b(u)r^{C_{GR}} + \frac{g_b(u)}{C_{GR}} = 0,
\end{eqnarray}
where $C_{GR}=2k-1$. For the same initial density profile and with identical time-dependence in the collapse dynamics (i.e. with identical forms of the functions $f_b(u)$ and $g_b(u)$),  the causal structure of the singularity in the brane world will be the same as that given by the standard Einstein equations when $C_{GR}=C_{BW}$, i.e. when
\begin{eqnarray} \label{k}
k_{BW}=\frac{1}{2}(k_{GR}-1)
\end{eqnarray}
where the subscripts BW and GR again refer to the values of $k$ in the brane world and in general relativity, respectively. From this point on these subscripts will be used whenever it is necessary to emphasize that we are referring to the value of $k$ in a specific theory, but will be dispensed with whenever such emphasis is unnecessary. However, since in both cases the parameter $k$ is limited to the range $k \in [0,1]$, we see that
\begin{eqnarray}
-1 \leq C_{GR} \leq 1,
1 \leq C_{BW} \leq 5.
\end{eqnarray}
Equivalence therefore occurs only for the specific value $C_{GR}=C_{BW}=1$, when $k_{GR}=1$ and $k_{BW}=0$, which corresponds to the case of the charged Vaidya metric discussed above. It is also clear that a whole range of possible solutions, corresponding to $-1 \leq C_{GR} < 0$, (i.e. $0 \leq k_{GR} < 1/2$), have no analogue on the brane. Physically however, such solutions are very interesting, since they represent the evolution of either flat space or of a naked singularity into a black hole embedded in a cosmology \cite{Hu96}. By contrast, all metrics of the form given in Eq. (\ref{baro_metric}) are asymptotically flat, for all possible values of $k$. A metric is asymptotically flat if its components obey the following relation as $r \rightarrow \infty$,
\begin{equation}
g_{\mu \nu} \rightarrow \eta_{\mu \nu} + \frac{\alpha_{\mu \nu}(x^c/r,t)}{r} + O\left(\frac{1}{r^{1+\epsilon}}\right)
\end{equation}
where $\alpha_{\mu \nu}$ is an arbitrary symmetric tensor, $x^c$ is a flat coordinate system at radial infinity, $\epsilon > 0$ is a constant and $\eta_{\mu \nu}$ is the Minkowski metric. For the metric in Eq. (\ref{baro_metric}) this requires that $C_{BW} \geq 0$, which is true for all $k_{BW}$ in the range $0 \leq k_{BW} \leq 1$. Likewise, a whole range of solutions corresponding to $1 < C_{BW} \leq 5$, (i.e. to all values of $k_{BW}$ not equal to zero), are inaccessible in general relativity.

In general, the position of any horizons, in either model, at a given time $u$, will be given by the positive, real solutions of the equation
\begin{eqnarray} \label{Horizon_Eqn}
r^{C+1}-2r^C f_b(u) + \frac{2g_b(u)}{C} = 0,
\end{eqnarray}
where $C=C_{GR}$ or $C=C_{BW}$. We may now investigate the properties of specific solutions within the class of spherically-symmetric collapse models by choosing two functions $f_b(u)$ and $g_b(u)$ which satisfy either the WEC or the DEC, (though if the DEC are satisfied, the WEC are satisfied automatically). One interesting choice of functions which satisfy the DEC; that is, for which $g_b(u) \geq 0$, $\partial g_b(u)/\partial u < 0$, and $\partial f_b(u)/\partial u > 0$ for all $u$, originally suggested by Hussain in the general-relativistic case \cite{Hu96}, is
\begin{eqnarray} \label{baro_metric_fns}
f_b(u) = \frac{1}{2}A\left[1 + \tanh(u)\right], \\
g_b(u) = \frac{1}{2}\left[1 - B\tanh(u)\right],
\end{eqnarray}
where $A$ and $B$ are constants such that $A \geq 0$ and $0 \leq B \leq 1$. With this choice, Eq. (\ref{Horizon_Eqn}) reduces to
\begin{eqnarray} \label{Horizon_Eqn*}
r^{C+1} = -\frac{(1+B)}{C},
\end{eqnarray}
for $u \rightarrow -\infty$, so that, for $-1 < C < 0$, there are no real solutions and there exists a naked singularity at past null infinity. For $u \rightarrow \infty$, the position of the horizons are given by the positive real solutions of
\begin{eqnarray} \label{Horizon_Eqn**}
r^{C+1} - 2Ar^C + \frac{(1+B)}{C} =0.
\end{eqnarray}
Eqs. (\ref{Horizon_Eqn*})-(\ref{Horizon_Eqn**}) recover the results obtained by Hussain in \cite{Hu96} when $C=C_{GR}= 2k_{GR}-1$. He showed that, for values of $k_{GR}$ in the range $1/2 < k_{GR} < 1$, there exists a naked singularity at $r=0$ in the limit $u \rightarrow -\infty$ but that, in the limit $u \rightarrow \infty$, there may exist horizons depending on the relative values of $A$ and $B$. For $k_{GR}=1$ these appear at $r=A \pm \sqrt{A^2 + B-1}$, which are the solutions of Eq. (\ref{Horizon_Eqn**}) for $C=1$. The key physical feature of this solution is that, for $A \neq 0$, a black hole with nonzero mass first forms when $A^2=1-B$, creating a mass gap between black hole and naked singularity solutions which contrasts with critical behavior solutions found in previous work \cite{Or90,Ch93,Ab93,Co94}.  In accordance with Eq. (\ref{k}), we confirm the existence of identical solutions (within the limit of the approximations taken here), in the brane world for $k_{BW}=0$. Crucially, whilst similar solutions exist for $k_{GR}>1/2$ but not for $k_{GR}<1/2$ (or for the critical case of $k_{GR}=1/2)$, in general relativity, asymptotically flat space-times are obtained in the brane world for all values of $k_{BW}$. Specifically, setting $k_{BW}=0$, $A=0$ and $B=1$ describes the evolution of a naked singularity at $u \rightarrow -\infty$ into flat space at  $u \rightarrow \infty$, which mirrors the corresponding result for $k_{GR}=1$ given in \cite{Hu96}.

Another feature that the metric, Eq. (\ref{baro_metric}) has in common with the equivalent general-relativistic case is that it permits the existence of black holes with null fluid hair. In the standard scenario, choosing metric functions of the of the form given in Eq.(\ref{baro_metric_fns}), which allow the DEC to be satisfied, lead to the existence hairy black holes at $u \rightarrow \infty$ for $k_{GR}$ in the range $1/2<k_{GR}<1$. This is true for any other choice of $f_b(u)$ and $g_b(u)$ that reach nonzero limiting values as $u \rightarrow \infty$ and these metrics ``lie between" the Schwarzschild and Reissner-Nordstrom solutions in the sense that the exponent of the second $r$-dependent term in $g_{rr}$, $-(C_{GR}+1)=-2k_{GR}$, lies between $-1$ and $-2$. However, as discussed above, for similar choices of  $f_b(u)$ and $g_b(u)$ in the metric, Eq. (\ref{baro_metric_fns}), a Reissner-Nordstrom-type solution is obtained at $u \rightarrow \infty$ only for $k_{BW}=0$, ($1-B=Q^2$). Whilst values of $k_{BW}$ in the range $0 <k_{BW} \leq 1$ also give rise to black holes with null fluid hair therefore, these solutions automatically ``exceed" the Reissner-Nordstrom solution in that the exponent of the second $r$-dependent term in $g_{rr}$, $-(C_{BW}+1)=-(4k_{GR}+2)$ is less than $-2$. It is interesting that, both in general relativity and in the brane world model, arguably the most realistic scenario for the gravitational collapse of compact objects in the early universe, the collapse of radiation with $p=(1/3) \rho$, leads generically to the existence of hairy black holes at future null infinity for reasonable choices of the functions characterizing the initial distribution and injection of the null fluid.  However, as shown above, the horizon structure differs considerably between the two cases.

\section{Solution of the field equations for the polytropic equation of state} \label{Section3.2}

We now consider the case where the energy density $\rho(u,r)$ and the pressure $p(u,r)$ are related by the polytropic equation of state,
\begin{equation} \label{poly}
p = k \rho^a,
\end{equation}
where $0 \leq k \leq 1$ and $a>0$, ($a \neq 1$). Whereas $k$ is a dimensionless constant in the barotropic equation of state, it now has units of $[l]^{4(a-1)}$. Substituting for $p$ from Eq. (\ref{poly}) into Eq. (\ref{presf}) and neglecting linear terms in $\rho$, then yields
\begin{equation} \label{presB}
-\frac{1}{r}\frac{\partial^2m}{\partial r^2} = \alpha \rho^2 + 2\alpha k \rho^{1 + a}
\end{equation}
so that substituting for $\rho$ from Eq. (\ref{dens_A}) gives
\begin{equation} \label{fieldeqB}
\frac{\partial^2m}{\partial r^2} + \frac{2}{r}\frac{\partial m}{\partial r}  + \frac{2\beta k}{r^a}\left(\frac{\partial m}{\partial r} \right)^{\frac{1+a}{2}}=0,
\end{equation}
where
\begin{equation} \label{beta}
\beta = \alpha\left(\frac{2}{\alpha}\right)^{\frac{1+a}{2}}.
\end{equation}

The solution then proceeds as follows; making the change of variables $z(u,r) = (\partial m/\partial r)^{\frac{1-a}{2}}$ allows Eq. $(\ref{fieldeqB})$ to be rewritten as
\begin{equation} \label{zEOM_B}
\frac{\partial z}{\partial r} + \frac{(1-a)}{r}z + \frac{(1-a)\beta k}{r^a}  = 0.
\end{equation}
Now, since $z(u,r)=C(u)r^{a-1}$, where $C(u)$ is an arbitrary function of $u$, is a solution of the homogenous equation $\frac{\partial z}{\partial r} + \frac{(1-a)}{r}z=0$, we may use the method of the variation of constants to search for a solution of the form
\begin{eqnarray} \label{var_const}
z(u,r) = C(u,r) r^{a-1},
\end{eqnarray}
to the inhomogenous equation. Substituting Eq. (\ref{var_const}) back into Eq. (\ref{zEOM_B}) and integrating with respect to $r$ gives
\begin{eqnarray}
C(u,r) = -\frac{1}{2}\beta k r^{-2a+2} + g_p(u),
\end{eqnarray}
so that the general solution for $z(u,r)$ is
\begin{eqnarray} \label{zB}
z(u,r) =  - \frac{1}{2}\beta k r^{1-a} + g_p(v)r^{-(1-a)} .
\end{eqnarray}
Finally,
\begin{eqnarray} \label{m'B}
\frac{\partial m}{\partial r} =  \left[ g_p(u)r^{-(1-a)} - \frac{1}{2}\beta k r^{1-a}\right]^{\frac{2}{1-a}}
\end{eqnarray}
so that
\begin{eqnarray} \label{mB}
m(u,r) = f_p(u) +  \int \frac{dr}{r^2} \left[ g_p(u)-\frac{1}{2}\beta k r^{2(1-a)}\right]^{\frac{2}{1-a}}.
\end{eqnarray}
Then, rewriting $\beta$ using Eq. (\ref{beta}) gives
\bea \label{mB*}
m(u,r) &=& f_p(u) +  \nonumber\\
&&\int \frac{dr}{r^2} \left[ g_p(u)-\left(\frac{\alpha}{2}\right)^{\frac{1-a}{2}} k r^{2(1-a)}\right]^{\frac{2}{1-a}},
\eea
where we recall that $\alpha = 1/(2\lambda_b) = k_5^4/(12k_4^2)$ by Eq. (\ref{alpha}). This solution may be compared with the equivalent one obtained in general relativity and quoted in \cite{Hu96} (here converted to the units and notation adopted in this paper);
\bea \label{mB_GR}
m(u,r) &=& f_p(u) + \nonumber\\
&&\int dr \left[g_p(u) - \left(\frac{k_4^2}{2}\right)^{1-a}k r^{2(1-a)}\right]^{\frac{1}{1-a}}.
\eea
Considering appropriate energy conditions, we see immediately that the WEC and first condition of the DEC are automatically satisfied for the polytropic equation of state for $\rho(u,r) \geq 0$, ($\partial m/\partial r \geq 0$), which requires
\begin{eqnarray} \label{ineq1}
g_p(u) \geq \left(\frac{\alpha}{2}\right)^{\frac{1-a}{2}} k r^{2(1-a)},
\end{eqnarray}
and that the second condition of the DEC gives $\rho^{2-a}(u,r) \geq k/\alpha$, requiring
\begin{eqnarray} \label{ineq2}
g_p(u) \geq \frac{1}{2}\alpha \left(\frac{k}{\alpha}\right)^{\frac{1}{2-a}} r^{4k+4}.
\end{eqnarray}

Which of these is the most stringent constraint will depend on the precise values of $k$ and $a$ but, in principle, both can be satisfied simultaneously. As with the barotropic fluid, we may again consider the maximum radius of the collapsing sphere, $R(u)$, and choose this function so as to satisfy the inequalities in Eqs. (\ref{ineq1})-(\ref{ineq2}) so that, in effect, $g_p(u)$ characterizes the maximal radius of the collapsing polytropic fluid. Since the final condition reduces to $\mu^{eff}(u,r)\geq 0$, or equivalently $\partial m/\partial u \geq 0$, this requires
\bea
&&\frac{df_p(u)}{d u} +\nonumber\\
&&\frac{2}{1-a}\frac{dg_p(u)}{d u}\int \frac{dr}{r^2} \left[ g_p(u)-\left(\frac{\alpha}{2}\right)^{\frac{1-a}{2}} k r^{2(1-a)}\right]^{\frac{1+a}{1-a}} \geq 0,\nonumber\\
\eea
which is also closely analogous to the equivalent condition in GR,
\bea
&&\frac{df_p(u)}{d u} + \nonumber\\
&&\frac{1}{1-a}\frac{dg_p(u)}{d u}\int dr \left[ g_p(u)-\left(\frac{k_4}{2}\right)^{1-a} k r^{2(1-a)}\right]^{\frac{a}{1-a}} \geq 0.\nonumber\\
\eea

For  $a<1$, these constraints can, in principle, be satisfied if either $df_p(u)/du \leq 0$ or $dg_p(u)/du \leq 0$ for some values of $u$, but by far the simplest way to satisfy them is to set  $dg_p(u)/du \geq 0$, $dg_p(u)/du \geq 0$ for all $u$, in both general relativity and in the brane world. In the former, imposing $a<1$ to satisfy the DEC leads to cosmological metrics, whereas imposing only the WEC allows $a>1$ but gives asymptotically flat metrics \cite{Hu96}. In the latter however, although the allowed values of $a$ are the same when imposing either the DEC or the WEC, the additional factor of $r^{-2}$ multiplying the terms inside the square brackets in the integral ensures that the resulting metrics are asymptotically flat, regardless of how large $a$ becomes. As for the collapse of a null fluid governed by the barotropic equation of state therefore, the brane world scenario seems to favor the formation of asymptotically flat metrics, in accordance with our physical intuition regarding the effect of increased effective pressure.

\section{Solution of the field equations for free strange quark matter} \label{Section3.3}

An equation of state for deconfined quark matter may be obtained from perturbation theory in QCD. Neglecting quark masses in the first order perturbation, the relation between pressure and energy density is given by
\begin{equation} \label{bag}
p(u,r) = \frac{1}{3}\left[\rho(u,r) - 4B \right],
\end{equation}
where $B \approx 57 MeV$ fm$^{-3} \approx 10^{14}$ gcm$^{-3}$ is the difference in energy density between the perturbative and the non-perturbative QCD vacuums. This model is known as the MIT bag model and the constant $B$ is called the bag constant. The collapse of a null quark fluid is of special interest to astrophysics as it is predicted that the temperature and pressure in the cores of some neutron stars is sufficient to induce a neutron-quark matter phase transition. The most stable form of matter in these cores is expected to be a plasma of deconfined strange quarks and the metric corresponding to the collapse of a null fluid governed by Eq. (\ref{bag}) provides a model for the collapse of a neutron star core. We begin again by substituting for $p(u,r)$ into Eq. (\ref{presf}) from our equation of state. However, the density of strange matter in the core is expected to be of order $\rho \approx 5 \times 10^{14}$ gcm$^{-3}$, which is comparable to the magnitude of the bag constant. We must therefore be careful to keep both $\alpha \rho^2(u,r)$ and $B \rho(u,r)$ terms in the high density limit and neglect only the other $\rho(u,r)$ terms in Eq. (\ref{presf}), yielding
\begin{equation} \label{presC}
-\frac{1}{r}\frac{\partial^2 m}{\partial r^2} = \alpha \rho^2 + \frac{2}{3} \alpha \left(\rho - 4B \right)\rho.
\end{equation}
Substituting for $\rho(u,r)$ from Eq. (\ref{dens_A}) and rearranging gives
\begin{equation}
\frac{\partial^2 m}{\partial r^2}  + \frac{10}{3r}\frac{\partial m}{\partial r}  - \frac{8}{3} \sqrt{2\alpha} B \sqrt{\frac{\partial m}{\partial r}} = 0.
\end{equation}
Using the change of variables $y(u,r)=\sqrt{\partial m/\partial r}/r$, this equation may be rewritten in the form
\begin{eqnarray}
\frac{\partial y}{\partial r} + \eta\frac{y}{r} - \frac{\gamma}{r} = 0,
\end{eqnarray}
where $\eta = 8/3$ and $\gamma = 4\sqrt{2\alpha}B/3$. Making the further substitution $\Theta(r)=\ln(r)$ allows a separation of variables so that
\begin{eqnarray}
\int \frac{dy}{\gamma-\eta y} = \int d\theta,
\end{eqnarray}
and
\begin{eqnarray}
y(u,r) =  \frac{\gamma}{\eta} + g_s(u) r^{-\eta}.
\end{eqnarray}
We then have
\begin{eqnarray}
\frac{\partial m}{\partial r} = \left(\frac{\gamma}{\eta}r + g_s(u) r^{1-\eta} \right)^2,
\end{eqnarray}
and
\bea
m(u,r) &=& f_s(u) + \frac{1}{3}\left(\frac{\gamma}{\eta}\right)^2r^3 + \nonumber\\
&&\frac{2}{(3-\eta)}\frac{\gamma}{\eta} g_s(u)r^{3-\eta} + \frac{g_s^2(u)}{(3-2\eta)}r^{3-2\eta},
\eea
so that, including the explicit values of $\eta$ and $\gamma$, the expression for the mass profile in terms of the constants $\alpha$ and $B$ is;
\begin{eqnarray}
m(u,r) = f_s(u) + \frac{1}{6}\alpha B^2 r^3 + 6\sqrt{\frac{\alpha}{2}}B g_s(u)r^{\frac{1}{3}} - \frac{3}{7}\frac{g_s^2(u)}{r^{\frac{7}{3}}}.\nonumber\\
\end{eqnarray}
Finally, substituting for $\alpha = 1/(2\lambda_b)$ and absorbing a factor of $3B/\sqrt{\lambda_b}$ into the definition of $g_s(u)$ gives
\begin{eqnarray} \label{baro}
m(u,r) &=& f_s(u) + \frac{1}{6}B\left(\frac{B}{2\lambda_b}\right) r^3 +\nonumber\\
&&g_s(u)r^{\frac{1}{3}} - \frac{1}{21}\frac{\lambda_b}{B^2}\frac{g_s^2(u)}{r^{\frac{7}{3}}}.
\end{eqnarray}
This may be compared with the corresponding expression for the mass profile of a collapsing mixture of strange quark fluid and radiation in general relativity, obtained by Harko and Cheng \cite{HaCh00},
\begin{eqnarray}
m(u,r) = f_s(u) + \frac{4\pi}{3} B r^3 +  g_s(u)r^{\frac{1}{3}} - \frac{q^2(u)}{2r},
\end{eqnarray}
where the function $q(u)$ characterizes the evolution of the vector potential, $A_{\mu}(u,r) = (q(u)/r) \delta^u{}_{\mu}$. Ignoring factors of $8\pi$ etc, which depend on the conventional choice of units, we see that, for $q(u)=0$, the two expressions differ in regard to the dimensionless multiplying factor of the $B r^{3}$ term and via the presence of an additional term in the brane world, caused by the $\rho^2(u,r)$ contributions to the effective energy-momentum tensor on the brane. Whilst a term proportional to $B r^3$ is present in the mass profile in both cases,  in the general-relativistic solution its multiplying factor is of order unity, whereas, in the brane world, its magnitude in characterized by the ratio
\begin{eqnarray} \label{Gamma}
\Gamma = \frac{B}{2\lambda_b}.
\end{eqnarray}
For $\Gamma \sim \mathcal{O}(1)$, ($\lambda_b \sim B$) the dynamics and profile of strange quark fluid in the two scenarios is similar for large $r$, given equivalent choices of $f_s(u)$ and $g_s(u)$, but the evolution close to the centre of the collapsing region is profoundly modified by the brane world  corrections to the gravitational field equations. This solution is consistent with our earlier results in the sense that, setting $B=0$ in Eq. (\ref{baro}) and identifying
\begin{eqnarray}
g_s^2(u) \leftrightarrow g_b(u),
\end{eqnarray}
recovers the solution for the barotropic equation of state with $k=1/3$. Written explicitly, the fluid flow along the outgoing, radial, null geodesics, the effective energy-density and the effective pressure are;
\begin{eqnarray} \label{beg}
\mu^{eff}(u,r) &=& \frac{2}{r^2}\Bigg[\frac{df_s(u)}{du} + \frac{dg_s(u)}{du}r^{\frac{1}{3}} - \nonumber\\
&&\frac{2}{21}g_s(u) \frac{dg_s(u)}{du}r^{-\frac{7}{3}}\Bigg], \nonumber\\
\end{eqnarray}
\begin{eqnarray}
\rho^{eff}(u,r) &=& \frac{2}{r^2}\Bigg[\frac{1}{2}B\left(\frac{B}{2\lambda_b}\right) r^2 + \frac{1}{3}g_s(u)r^{-\frac{2}{3}} + \nonumber\\
&&\frac{1}{9}\frac{\lambda_b}{B^2}g_s^2(u)r^{-\frac{10}{3}}\Bigg],
\end{eqnarray}
\begin{eqnarray} \label{end}
p^{eff}(u,r)& =& \frac{2}{r^2}\Bigg[-\frac{1}{2}B\left(\frac{B}{2\lambda_b}\right) r^2 + \frac{1}{9}g_s(u)r^{-\frac{2}{3}} + \nonumber\\
&&\frac{5}{27}\frac{\lambda_b}{B^2}g_s^2(u)r^{-\frac{10}{3}}\Bigg],
\end{eqnarray}
which may be compared with their counterparts in general relativity, obtained in \cite{HaCh00}
\begin{eqnarray}
\mu(u,r) = \frac{1}{4\pi r^2}\left[\frac{df_s(u)}{du} + \frac{dg_s(u)}{du}r^{\frac{1}{3}} -\frac{q(u)}{r}\frac{dq(u)}{du}\right],
\end{eqnarray}
\begin{eqnarray}
\rho(u,r) = \frac{1}{4\pi r^2}\left[4\pi Br^2 + \frac{1}{3}g_s(u)r^{-\frac{2}{3}} + \frac{q^2(u)}{2r^2}\right],
\end{eqnarray}
\begin{eqnarray}
p(u,r) = \frac{1}{12\pi r^2}\left[-12\pi B r^2 + \frac{1}{3}g_s(u)r^{-\frac{2}{3}} + \frac{q^2(u)}{2r^2}\right],
\end{eqnarray}
by setting $q(u)=0$. Alternatively, we note that the additional terms coming from the  $\rho^2(u,r)$ corrections to the gravitational field equations on the brane play a similar role to the gauge field terms for $q(u) \neq 0$ in the general-relativistic case. Roughly speaking, replacing $q(u) \neq 0$ in the expression for $m(u,r)$, $\rho(u,r)$, $p(u,r)$ or $\mu(u,r)$ in the latter with $(\lambda_b/B^2)g_s^2(u)r^{-\frac{4}{3}}$, yields the equivalent expression for  $m(u,r)$, $\rho^{eff}(u,r)$, $p^{eff}(u,r)$ or $\mu^{eff}(u,r)$ in the former, with $q(u)=0$. We now consider the energy conditions. The first component of either the WEC or the DEC, $p^{eff}(u,r) \geq 0$, requires
\begin{eqnarray} \label{WEC1}
r^{\frac{16}{3}} - \frac{2}{9}\left(\frac{2\lambda_b g_s(u)}{B^2}\right)r^{\frac{8}{3}} - \frac{5}{27}\left(\frac{2\lambda_b g_s(u)}{B^2}\right)^2 \leq 0.
\end{eqnarray}
Setting $X = r^{\frac{8}{3}}$ and $A(u) = 2\lambda_b g_s(u)/B^2$, this may be rewritten as a quadratic,
\begin{eqnarray}
X^2 - \frac{2}{9}A(u)X - \frac{5}{27}A^2(u) \leq 0,
\end{eqnarray}
from which we see that the condition in Eq. (\ref{WEC1}) holds within the region $X \in [-A(u)/3,5A(u)/9]$. Setting $g_s(u) \geq 0$, so that $A(u) \geq 0$ for all $u$, then,  if the strange quark fluid initially extends from $r=0$ to some maximal radius $R$, the condition $p^{eff}(u,r) \geq 0$ reduces to $A(u) \geq(3/5)R^{\frac{8}{3}}$ and, in general, if the time-dependent radius of the collapsing fluid is given by $R(u)$, we require
\begin{eqnarray} \label{ec}
A(u) \geq \frac{9}{5}R^{\frac{8}{3}}(u).
\end{eqnarray}
As before, strictly speaking, in this case we require an appropriate set of matching conditions in order to construct a regular solution for the region containing both the collapsing stellar core and the surrounding inter-stellar medium. Since, in this paper, we are interested mainly in the core collapse, we leave such a solution to a future publication, though its construction should not pose any major theoretical problems.
In principle, the first component of the energy conditions may also be satisfied if $g_s(u) \leq 0$ for all $u$. In this case, $-A(u) \leq 3 R^{\frac{8}{3}}(u)$ is required instead of Eq. (\ref{ec}) though, in either case, it may be seen that $g_s(u)$ characterizes the time-dependent radius of the collapsing fluid. The second component of the WEC, $\rho^{eff}(u,r) \geq 0$, reduces to
\begin{eqnarray}
X^2 + \frac{2}{3}A(u)X + \frac{1}{9}A^2(u) \leq 0,
\end{eqnarray}
which is satisfied for $X \geq -A(u)/3$. Therefore, setting $g_s(u) \geq 0$, ($A(u) \geq 0$), and adopting the constraint in Eq. (\ref{ec}) is the simplest way of satisfying the WEC for all $r$ in the range $0 \leq r \leq R(u)$. For $\rho^{eff}(u,r) \approx \alpha \rho^2(u,r) \geq 0$, the second component of the brane world DEC, $\rho^{eff}(u,r) \geq p(u,r)$, requires
\begin{eqnarray}
\alpha \rho^2 - \frac{1}{3}\rho + \frac{1}{3}B \geq 0,
\end{eqnarray}
which is satisfied for $\rho(u,r)$ in the range $\rho(u,r) \in [(\lambda_b/3)(1-\sqrt{1-6B/\lambda_b}),(\lambda_b/3)(1+\sqrt{1-6B/\lambda_b})]$. Assuming the reality of $\rho(u,r)$, this yields a condition on the functions $f_s(u)$ and $g_s(u)$, but it is interesting that the reality condition itself requires $\Gamma \leq 1/12$ or, equivalently,
\begin{eqnarray}
B \leq \frac{k_4^2}{k_5^4}.
\end{eqnarray}
However, we must be careful in considering the approximations we have made to obtain the solution, Eqs. (\ref{beg})-(\ref{end}). Adopting instead the (more accurate) approximation $\rho^{eff} \approx \rho(u,r) + \alpha \rho^2(u,r)$, the condition $\rho^{eff}(u,r) \geq p(u,r)$ reduces to
\begin{eqnarray}
\alpha \rho^2 + \frac{2}{3}\rho + \frac{1}{3}B \geq 0,
\end{eqnarray}
which is trivially satisfied for $\rho(u,r) \geq 0$. This is, in turn, is automatically satisfied  by the condition in Eq. (\ref{ec}), as long as we define $\rho(u,r) = +\sqrt{\rho^{eff}(u,r)/\alpha}$. Therefore, although the high-density approximation $\rho^{eff}(u,r) \approx  \alpha \rho^2(u,r)$, rather than $\rho^{eff}(u,r) \approx \rho(u,r) + \alpha \rho^2(u,r)$ was used to obtain Eqs. (\ref{beg})-(\ref{end}) and Eq. (\ref{ec}), it is still physically reasonable to assume that the second part of the DEC is trivially satisfied, given the fulfillment of the WEC, for a strange quark fluid in the brane world. Finally, the third component of the DEC,  $\mu^{eff}(u,r) \geq 0$, implies
\begin{eqnarray} \label{mu}
\frac{df_s(u)}{du} + \frac{dg_s(u)}{du}r^{\frac{1}{3}}  \geq \frac{2}{21}g_s(u) \frac{dg_s(u)}{du}r^{-\frac{7}{3}}.
\end{eqnarray}
For small values of $r$ the right-hand-side of Eq. (\ref{mu}) dominates and the DEC cannot hold, unless we assume that the function $g_s(u)$ behaves such that $dg_s^2(u)/du \rightarrow 0$ as $r \rightarrow 0$. In general relativity, a similar problem occurs for charged quark matter, where the condition $\mu(u,r)\geq 0$ reduces to
\begin{eqnarray} \label{mu}
\frac{df_s(u)}{du} + \frac{dg_s(u)}{du}r^{\frac{1}{3}}  \geq \frac{q(u)}{r}\frac{dq(u)}{du},
\end{eqnarray}
and it is necessary to assume that $dq^2(u)/du \rightarrow 0$ as $r \rightarrow 0$, though for $q(u) =0$, the DEC are easily satisfied by setting $df_s(u)/du \geq 0$ and $dg_s(u)/du \geq 0$ \cite{HaCh00}. Alternatively, in both scenarios, we may assume that, at small radii, matter is converted to strange quark matter so as to satisfy the final part of the DEC.

Having seen that both the WEC and the DEC may be satisfied for appropriate physical assumptions and choices of the arbitrary functions $f_s(u)$ and $g_s(u)$, we now wish to investigate the final state of the collapsing quark fluid. In particular, we will attempt to identify under what circumstances, if any, a naked singularity, rather than a black hole, can occur, as well as to identify the causal structure of black hole solutions and any null fluid ``hair" they may posses. Assuming that, for $u \rightarrow \infty$, the functions $f_s(u)$ and $g_s(u)$ reach finite limiting values;
\begin{eqnarray}
\lim_{u \rightarrow \infty}f_s(u) = f_s = {\rm const.}, \;\;\;\lim_{u \rightarrow \infty}g_s(u) = g_s = {\rm const.},\nonumber\\
\end{eqnarray}
the positions of any apparent horizons are given by the real, positive solutions of the equation $r = 2\lim_{u \rightarrow 0}m(u,r)$, i.e
\begin{eqnarray}
r =  2f_s + \frac{1}{3}B\left(\frac{B}{2\lambda_b}\right) r^3 +g_s r^{\frac{1}{3}} - \frac{2}{21}\frac{\lambda_b}{B^2}\frac{g_s^2}{r^{\frac{7}{3}}},
\end{eqnarray}
which, setting $\chi = r^{\frac{1}{3}}$, reduces to the following polynomial of order sixteen in $\chi$;
\begin{eqnarray}
&&(\Gamma B)^2\chi^{16} - 3(\Gamma B)\chi^{10} +3(\Gamma B)g_s \chi^{8} + \nonumber\\
&&(\Gamma B)f_s \chi^{7} - \frac{3}{21}g_s^2 = 0.
\end{eqnarray}
The physical nature of the central singularity can then be recognized by evaluating the curvature tensor, $R_{\mu\nu}R^{\mu\nu}$, given by
\begin{eqnarray}
R_{\mu\nu}R^{\mu\nu}  = \frac{4}{r^2}\left[B\Gamma + g_s(u) r^{-\frac{8}{3}} + \frac{2}{9}\frac{g_s^2(u)}{B\Gamma}r^{-\frac{10}{3}}\right],
\end{eqnarray}
which diverges as $r \rightarrow 0$. However, to determine under what (if any) circumstances the shell-focussing singularity is naked, we must investigate the outgoing, radial, null geodesics for specific choices of the arbitrary functions $f_s(u)$ and $g_s(u)$. In  the general-relativistic case, Harko and Cheng \cite{HaCh00} showed that setting
\be \label{fns}
f_s(u) =  \frac{\alpha_0 u}{2},  g_s(u) =  \frac{\beta_0 u^{\frac{2}{3}}}{2},  q(u) = \gamma_0 u,
\ee
with $\alpha_0 > 0$, $\beta_0 > 0$, and $\gamma_0 \geq 0$, respectively,
the geodesic equation, $dv^{\mu}/ds = \Gamma^{\mu}{}_{\nu\sigma}v^{\nu}v^{\sigma} = 0$, with $v^{\mu} = n^{\mu} = [-1,-(1/2)(1-2m/r),0,0]$, gives
\begin{eqnarray}   \label{geod}
\frac{du}{dr} = \frac{1}{1 - \alpha_0\left(\frac{u}{r}\right) - \beta_0\left(\frac{u}{r}\right)^{\frac{2}{3}} - \gamma_0\left(\frac{u}{r}\right)^2 - \frac{8\pi B}{3}r^2}.
\end{eqnarray}
For the geodesic tangent to be uniquely defined at the singular point $r=0$, $u=0$, the condition
\begin{eqnarray}  \label{lim}
X_0 = \lim_{u,r \rightarrow 0}\frac{u}{r} = \lim_{u,r \rightarrow 0}\frac{du}{dr},
\end{eqnarray}
must hold \cite{Jo93}. When the limit $X_0$ exists and is real and positive, the singularity is, at least locally, naked. For a geodesic equation of the form Eq. (\ref{geod}), the condition in Eq. (\ref{lim}) leads to the following algebraic equation
\begin{eqnarray} \label{roots}
\gamma_0 X_0^3 + \alpha_0 X_0^2 + \beta_0 X_0^{\frac{5}{3}} - X_0 +1 = 0,
\end{eqnarray}
which, setting $\chi_0=X_0^{\frac{1}{3}}$ and using Eq. (\ref{Gamma}), may be rewritten as a ninth order polynomial;
\begin{eqnarray}   \label{roots*}
f(\chi_0) = \gamma_0 \chi_0^9 + \alpha_0 \chi_0^6 + \beta_0 \chi_0^{5} - \chi_0^3 +1 = 0.
\end{eqnarray}

Likewise, in \cite{HaCh00} it was found that, in the general-relativistic case for $q(u) \neq 0$, the horizon equation reduced to a ninth order polynomial of the same form as that given in Eq. (\ref{roots*}), again highlighting the similarities between the presence of the brane world corrections and a charged fluid in four-dimensional general relativity. As also discussed in \cite{HaCh00}, due to a theorem by Poincar\'{e}, the number of positive roots of a polynomial equation $f(\chi_0)$ is equal to the number of changes in sign in the sequence of non-negative coefficients of the polynomial $g(\chi_0) = (1+\chi_0)^kf(\chi_0)$,  \cite{HNA} and, by this criterion, the polynomial in Eq. (\ref{roots*}) has two positive roots. A similar result can be obtained by using Descartes' rule of signs, according to which  the number of positive roots  of a single-variable polynomial with real coefficients  ordered by descending variable exponent is either equal to the number of sign differences between consecutive nonzero coefficients, or is less than it by an even number \cite{Des}. Therefore, in the general-relativistic case, it is always possible for the collapse of strange quark matter to lead to the existence of a (locally) naked singularity. In the brane world scenario, the same choice of functions, given in Eq. (\ref{fns}), (with $\gamma_0 = 0$), leads to
\begin{eqnarray}  
\frac{du}{dr} = \frac{1}{1 - \alpha_0\left(\frac{u}{r}\right) - \beta_0\left(\frac{u}{r}\right)^{\frac{2}{3}} - \frac{1}{3}B\Gamma r^2 + \frac{\beta_0^2}{84 B \Gamma}\left(\frac{u}{r}\right)^{\frac{4}{3}}\frac{1}{r^2}},\nonumber\\
\end{eqnarray}
so that the limit in Eq. (\ref{lim}) is not well defined. However, if instead we make the choice
\be \label{fns*}
f_s(u) =  \frac{\alpha_0 u}{2},  g_s(u) =  \frac{\beta_0 u^{\frac{5}{3}}}{2},
\alpha_0 > 0, \beta_0 > 0,
\ee
then
\begin{eqnarray}  
\frac{du}{dr} = \frac{1}{1 - \alpha_0\left(\frac{u}{r}\right) - \beta_0u\left(\frac{u}{r}\right)^{\frac{2}{3}} - \frac{1}{3}B\Gamma r^2 + \frac{\beta_0^2}{84 B \Gamma}\left(\frac{u}{r}\right)^{\frac{10}{3}}},\nonumber\\
\end{eqnarray}
which reduces to a thirteenth order polynomial in $\chi_0$;
\begin{eqnarray} \label{roots**}
\epsilon_0 \chi_0^{13} - \alpha_0 \chi_0^{6} + \chi_0^{3} - 1 = 0,
\end{eqnarray}
where $\epsilon_0 = \beta_0^2/84 B \Gamma$. Again, by the theorem quoted above, or by Descartes' rule of signs,  Eq.~(\ref{roots**}) has at least a single real, positive root.

In the brane world therefore, it is also possible for the collapse of a strange quark fluid to end in the formation of a locally naked singularity. However, for collapse ending in a black hole, the null fluid hair given by the causal structure of the horizons differs from that given by the collapse of a pure strange matter fluid (with equivalent initial conditions), in the general-relativistic case. Instead, it resembles the collapse of a neutral strange matter/charged matter mixture in the latter, (for a particular choice of $q(u)$). As in four-dimensional general relativity, the metrics produced by the spherically-symmetric collapse of strange matter in the brane world are cosmological.

\section{Solution of the field equations for a Hagedorn fluid} \label{Section3.4}

The Hagedorn equation of state can be used to model ordinary matter at very high densities such as those found in the early universe. As already mentioned above, in the Hagedorn model it is proposed that there exists an effective highest temperature for any system called the Hagedorn temperature $T_H$. This is based on the assumption that, at high densities, a large number of baryonic resonant states arise. In this case, increasing the pressure and energy density beyond the critical values corresponding to $T_H$, labelled $p_0$ and $\rho_0$, respectively, increases the number of particles but not the kinetic energy/temperature of the system. The equation of state is then
\begin{equation} \label{Hag}
p(u,r) = p_0 + \rho_0 \ln \left(\frac{\rho(u,r)}{\rho_0}\right).
\end{equation}
Empirically fitting the model to data, the values of the above parameters are estimated to be $T_H \approx 150-190$ MeV, $p_0 \approx 0.314 \times 10^{14}$ gcm$^{-3}$ and $\rho_0 \approx 1.253 \times 10^{14}$ gcm$^{-3}$ and it is thought that the equation of state, Eq. (\ref{Hag}) could hold for densities as high as ten times the nuclear density $\rho_n = 2 \times 10^{14}$ gcm$^{-3}$. Following the same type of procedure used in the previous section, i.e. substituting for $p(u,r)$ in Eq. (\ref{presf}) while neglecting linear terms of order $\rho(u,r)$, but keeping those in $\rho_0 \rho(u,r)$ and $p_0 \rho(u,r)$, we have
\begin{equation}
-\frac{1}{r}\frac{\partial^2 m}{\partial r^2} =  \alpha \rho^2 + 2\alpha p_0 \rho + 2\alpha \rho_0 \rho \ln\left(\frac{\rho}{\rho_0}\right),
\end{equation}
which, substituting for $\rho(u,r)$ from Eq. (\ref{dens_A}), yields
\bea \label{EOM_D}
&&\frac{\partial^2 m}{\partial r^2} + \frac{2}{r}\frac{\partial m}{\partial r} + 2\sqrt{2\alpha} \Bigg[p_0 + \nonumber\\
&&\rho_0 \ln \left(\frac{1}{\rho_0 r}\sqrt{\frac{2}{\alpha}}\sqrt{\frac{\partial m}{\partial r}} \right) \Bigg] \sqrt{\frac{\partial m}{\partial r}} = 0.
\eea
We note that the condition $\rho(u,r) \geq 0$ implies $\partial m/\partial r \geq 0$ and that we must take the positive square root, yielding the equation above. Again making the change of variables $y(u,r)=\sqrt{\partial m/\partial r }/r$, Eq. (\ref{EOM_D}) may be rewritten as
\begin{equation} \label{x}
\frac{\partial y}{\partial r} + \frac{2y}{r} + \frac{\sqrt{2\alpha}}{r} \left[p_0 + \rho_0 \ln \left(\sqrt{\frac{2}{\alpha}} \frac{y}{\rho_0}\right)\right] = 0,
\end{equation}
and, again using the substitution $\theta(r)=\ln(r)$, and defining
\begin{equation}
z(u,r) = \frac{\rho(u,r)}{\rho_0} =\sqrt{\frac{2}{\alpha}} \frac{y(u,r)}{\rho_0},
\end{equation}
we may perform a separation of variables so that
\begin{equation} \label{x_int}
\frac{1}{2}\int d\theta = - \frac{1}{2}\int \frac{dz}{z + 2 \ln(z) + 2 q},
\end{equation}
where
\begin{equation}
q = \frac{p_0}{\rho_0} \approx \frac{1}{4}.
\end{equation}
Finally, defining
\begin{equation}
w(u,r) = \ln \left(\frac{\rho(u,r)}{\rho_0}\right) = \ln \left(z(u,r)\right),
\end{equation}
gives
\begin{equation} \label{w_int}
\frac{1}{2}\int d\theta = -\frac{1}{2}\int \frac{e^{w}}{e^{w} + 2 w + 2 q}dw,
\end{equation}
and performing the integral on the left-hand-side yields
\bea \label{ratio}
\frac{r}{g_H(u)} &=& \exp \left(-\frac{1}{2}\int \frac{e^{w}}{e^{w} + 2 w + 2 q}dw\right) = \nonumber\\
&&\exp \left[-\frac{1}{2}F(w)\right].
\eea
Returning to the gravitational field equation for $\rho^{eff}(u,r)$,  Eq. (\ref{dens_A}) we have,
\begin{equation}
\frac{2}{r^2}\frac{\partial m}{\partial r} = \alpha \rho_0^2 e^{2w},
\end{equation}
so that the final solution for $m(u,r)$ is;
\begin{equation} \label{m_H}
m(u,r) = f_{H}(u) - \frac{\delta}{2} g_{H}^3(u) K(w),
\end{equation}
where we have defined
\begin{equation}
K(w) = \int \frac{\exp \left[3w - \frac{3}{2}F(w)\right]}{e^{w} + 2w + 2q}dw,
\end{equation}
and
\begin{equation}
\delta = \frac{\alpha \rho_0^2}{2}.
\end{equation}
This may be compared with the analogous solution obtained in general relativity \cite{Ha03},
\begin{equation}
m_{GR}(u,r) = f_{H}(u) - \frac{\rho_0}{2}g^3_{H}(u) K_{GR}(w),
\end{equation}
where
\begin{equation} \label{F_{GR}}
F_{GR}(w) = \int \frac{e^{w}}{e^{w} + w + q}dw.
\end{equation}
and
\begin{equation} \label{K_{GR}}
K_{GR}(w) = \int \frac{\exp \left[3w - \frac{3}{2}F_{GR}(w)\right]}{e^{w} + w + q}dw.
\end{equation}
The full solution in brane world is then given by Eq.~(\ref{m_H}), plus
\begin{equation} \label{soln2}
\rho^{eff}(w) = 2\delta e^{2w},
\end{equation}
\begin{equation} \label{soln3}
p^{eff}(w) =  2\delta e^{2w}(e^{w} + 2w + 2q),
\end{equation}
and
\bea \label{soln4}
\mu^{eff}(u,r)& =& \frac{2}{g_{H}^2(u)}\frac{df_{H}(u)}{du}e^{F(w)} - \nonumber\\
&&3\delta \frac{dg_{H}(u)}{du}e^{F(w)}K(w) + \nonumber\\
&&\delta \frac{dg_{H}(u)}{du} \frac{\exp \left[3w - F(w)\right]}{e^{w} + 2w +2q} \frac{dH}{d\eta},
\eea
where we have defined $\eta(u,r) = r/g_{H}(u)$ and $w = H(\eta)$. The functions $F(w)$ and $K(w)$ and the ratio $\eta(u,r) = r/g_{H}(u)$ are all expandable as power series in $w=w(u,r)$, giving
\bea \label{F_exp}
F(w) &=& \int \frac{e^{w}}{e^{w} + 2 w + 2 q}dw \approx \nonumber\\
&& \frac{w}{1+2q} + \frac{2(-1+q)}{2(1+2q)^2}w^2 + \mathcal{O}(w^3) + . \ . \ .
\eea
\bea \label{K_exp}
K(w) &=& \int \frac{\exp \left(3w - \frac{3}{2}F(w)\right)}{e^{w} + 2 w + 2 q}dw \approx \frac{w}{e^{\frac{3}{2}}(1+2q)} + \nonumber\\
&&\frac{11 + 8(-1+3q)}{8e^{\frac{3}{2}}(1+2q)^2}w^2 + \mathcal{O}(w^3) + . \ . \ .
\eea
\bea \label{r_exp}
\frac{r}{g_H(u)} &=& \exp \left(-\frac{1}{2}F(w)\right) \approx 1 - \frac{w}{2(1+2q)} + \nonumber\\
&&\frac{5-4q}{8(1+2q)^2}w^2 +  \mathcal{O}(w^3) + . \ . \ .
\eea
From Eq. (\ref{r_exp}) it follows that,
\begin{equation} \label{X}
r \frac{\partial w}{\partial r} = -2\frac{(e^w +2w + 2q)}{e^w},
\end{equation}
so that, in the limit of very high densities, $\rho \rightarrow \infty$, ($w \rightarrow \infty$),
\begin{equation}
w(u,r) \approx \ln \left(\frac{G_{H}(u)}{r^2}\right),
\end{equation}
where $G_{H}(u)$ is an arbitrary function of $u$ with dimensions $[l]^2$,
and the solution takes an exceptionally simply form;
\begin{equation}  \label{BW1}
m(u,r) \approx F_{H}(u) - \delta \frac{G_{H}^2(u)}{r},
\end{equation}
\begin{equation} \label{BW2}
\rho^{eff}(u,r) \approx 2\delta \frac{G_{H}^2(u)}{r^4} \approx 2\delta \frac{G_{H}(u)}{r^2},
\end{equation}
\begin{equation}  \label{BW3}
p^{eff}(u,r) \approx 2\delta \frac{G_{H}^2(u)}{r^4} + 4\delta \frac{G_{H}(u)}{r^2} \left[q + \ln \left(\frac{G_{H}(u)}{r^2}\right)\right],
\end{equation}
\begin{equation} \label{BW4}
\mu^{eff}(u,r) \approx \frac{2}{r^2}\frac{dF_{H}(u)}{du} -  \frac{4\delta}{r^3} G_{H}(u)\frac{dG_{H}(u)}{du},
\end{equation}
where we have used $G_{H}(u)/r^2 \approx e^{w} \approx 1 + w(u,r) + \mathcal{O}(w^2(u,r))$ in Eq. (\ref{BW2}). Comparing this to the solution obtained in general relativity in the $\rho \rightarrow \infty$ limit, Eqs.(\ref{GR1})-(\ref{GR4}), we see that the higher order corrections to the effective density and pressure in the brane world significantly affect the $r$-dependence of the mass profile for a collapsing Hagedorn fluid. From \cite{Ha03};
\begin{equation} \label{GR1}
m_{GR}(u,r) \approx F_{H}(u) + \rho_0 G_{H}(u) r - \frac{q^2(u)}{r},
\end{equation}
\begin{equation} \label{GR2}
\rho(u,r) \approx \rho_0 \frac{G_{H}(u)}{r^2},
\end{equation}
\begin{equation} \label{GR3}
p(u,r) \approx \rho_0 \left[q + \ln \left(\frac{G_{H}(u)}{r^2}\right)\right],
\end{equation}
\begin{equation} \label{GR4}
\mu(u,r) \approx \frac{1}{r^2}\frac{dF_{H}(u)}{du} +  \frac{2\rho_0}{r}\frac{dG_{H}(u)}{du} - \frac{2}{r^3} q(u) \frac{dq(u)}{du}.
\end{equation}
Interestingly, as in the case of the strange quark fluid, we again see that the additional terms generated by corrections to the gravitational field equations on the brane seem to play a similar role to the additional terms generated by the presence of an electromagnetic field, with vector potential $A_{\mu}=(q(u)/r)\delta^{u}{}_{\mu}$, in the general-relativistic case. This time the quantity $\alpha \rho_0^2 G_H(u)$ in the expressions for $m(u,r)$, $\rho^{eff}(u,r)$, $p^{eff}(u,r)$ and $\mu^{eff}(u,r)$ in Eqs. (\ref{BW1})-(\ref{BW4}) plays {\it almost} the same role as $\rho_0 q(u)$ in the expressions for $m(u,r)$, $\rho(u,r)$, $p(u,r)$ and $\mu(u,r)$ in Eqs. (\ref{GR1})-(\ref{GR4}), though the expressions for $p^{eff}(u,r)$ and $\mu^{eff}(u,r)$ are not identical to those for $p(u,r)$ and $\mu(u,r)$ under this correspondence due to the differing number of terms contained in each.

Since it is appropriate to take the $\rho \rightarrow \infty$ limit for $r \rightarrow 0$, the solution in Eqs. (\ref{BW1})-(\ref{BW4}) should be accurate close to the core of the collapsing fluid. In order to find the behavior of the general solution, given by Eq. (\ref{m_H}) together with Eqs. (\ref{soln2})-(\ref{soln4}), in the opposite limit, i.e. for large, but finite $r$, we also follow an analogous procedure to that outlined in \cite{Ha03}. We begin by noting that the boundary of the Hagedorn fluid is defined by the equation $p(u,r)=0$ or, equivalently $w(u,r) = -q$. Therefore, we define the critical value of $w(u,r)$, beyond which the solution becomes unphysical, as $w_s=-q$ and search for a solution of the form
\begin{equation} \label{w_s}
w(u,r) = w_s + w_1(u,r), \ \ \ |w_1(u,r)| << w_s,
\end{equation}
close to the boundary. Substituting Eq. (\ref{w_s}) into Eq. (\ref{X}), we obtain
\begin{equation}
r \frac{\partial w}{\partial r} = -2\frac{(2w_1 + e^{w_s}e^{w_1})}{e^{w_s}e^{w_1}}.
\end{equation}
Rewriting $s=e^{w_s}$, expanding $e^{w_1}$ to linear order and integrating, this gives
\bea \label{al}
&&\frac{s}{2+s}w_1 + \frac{2s}{(2+s)^2}\ln(s) +\nonumber\\
&& \frac{2s}{(2+s)^2} \ln \left[1 + s+
 \frac{(2+s)}{s}w_1\right] =
 \ln \left[\frac{G_{H}(u)}{r^2}\right]. \nonumber\\
\eea
Expanding the final logarithm of the left-hand-side of Eq. (\ref{al}), the approximate solution for $w(u,r)$ close to the boundary of the Hagedorn fluid is, simply
\begin{equation}
w(u,r) = w_0 + \ln \left[\frac{G_{H}(u)}{r^2}\right],
\end{equation}
where we have defined
\begin{equation}
w_0 = \frac{1-2s}{(2+s)^2}.
\end{equation}
The expressions for the physically relevant quantities are;
\begin{equation} \label{BW1*}
m(u,r) \approx F_{H}(u) - \delta \frac{G_{H}^2(u)}{r} e^{2w_0},
\end{equation}
\begin{equation} \label{BW2*}
\rho^{eff}(u,r) \approx 2\delta \frac{G_{H}^2(u)}{r^4} e^{2w_0},
\end{equation}
\bea  \label{BW3*}
p^{eff}(u,r) &\approx & 2\delta \frac{G_{H}^2(u)}{r^4} e^{2w_0} +\nonumber\\
 &&4\delta \frac{G_{H}(u)}{r^2} e^{w_0} \left[q+w_0 + \ln \left(\frac{G_{H}(u)}{r^2}\right)\right], \nonumber\\
\eea
\begin{equation} \label{BW4*}
\mu^{eff}(u,r) \approx \frac{2}{r^2}\frac{dF_{H}(u)}{du} -  \frac{4\delta}{r^3} G_{H}(u)\frac{dG_{H}(u)}{du} e^{w_0}.
\end{equation}
Again, we must now consider an appropriate set of energy conditions. The WEC, $p^{eff}(u,r) \geq 0$, $\rho^{eff}(u,r) \geq 0$, are trivially satisfied for all $r$ and $u$. The second component of the DEC,  $\rho^{eff}(u,r) \geq p(u,r)$, requires
\begin{equation}
G_{H}^2(u) \geq \frac{\rho_0}{\alpha}e^{-2w_0}r^4 \left[q + w_0 + \ln \left(\frac{G_{H}(u)}{r^2}\right)\right],
\end{equation}
and the third component, $\mu^{eff}(u,r) \geq 0$, yields
\begin{equation} \label{DEC_H}
\frac{d F_{H}(u)}{du} \geq 2\delta G_{H}(u) \frac{dG_{H}(u)}{du}\frac{e^{w_0}}{r},
\end{equation}
close to the boundary, while equivalent expressions with $w_0$ set equal to zero are valid close to the core. Yet again, the brane world corrections seem to mirror the contribution of an electromagnetic component in general relativity and Eq. (\ref{DEC_H}) closely resembles the equivalent result, obtained in \cite{Ha03}, for $q(u) \neq 0$;
\begin{equation}
\frac{d F_{H}(u)}{du} + \rho_0 \frac{d G_{H}(u)}{du} \geq 2 q(u) \frac{dq(u)}{du}\frac{1}{r},
\end{equation}
which requires matter to be converted close to $r=0$, unless $dq^2(u)/du \rightarrow 0$ for $r \rightarrow 0$. In the brane world, we must assume $dG_{H}^2(u)/du \rightarrow 0$ for $r \rightarrow 0$, or that matter is likewise converted near to the core of the collapsing region. We can now examine the structure of the apparent horizons and the singularity and determine the conditions under which naked singularities can occur for specific choices of $F_{H}(u)$ and $G_{H}(u)$. Assuming that these functions reach constant limiting values at future full infinity;
\begin{equation}
\lim_{u \rightarrow \infty} F_{H}(u) = F_{H}, \ \ \ \lim_{u \rightarrow \infty}G_{H}(u) = G_{H}
\end{equation}
the positions of the apparent horizons are given by the positive real solutions of the equation
\begin{equation} \label{H_Hor}
F_{H} = \frac{\delta}{2}G_{H}^3K(w) + G_{H}e^{-\frac{1}{2}F(w)},
\end{equation}
which, in general, will have multiple solutions. The expression in Eq. (\ref{H_Hor}) is completely analogous to that obtained in general relativity (though, this time, for $q(u)=0$ \cite{Ha03}), namely
\begin{equation} \label{H_Hor*}
F_{H} = \frac{\rho_0}{2}G_{H}^3K_{GR}(w) + G_{H}e^{-\frac{1}{2}F_{GR}(w)},
\end{equation}
where $F_{GR}(w)$ and $K_{GR}(w)$ are defined as in Eqs. (\ref{F_{GR}})-(\ref{K_{GR}}), above. The physical nature of the central singularity can again be recognized by evaluating the explicit expression for the curvature tensor $R_{\mu\nu}R^{\mu\nu}$,
\begin{eqnarray}
R_{\mu\nu}R^{\mu\nu}  = \frac{16 G_s^2(u)}{r^8},
\end{eqnarray}
which diverges for $r \rightarrow 0$.

For the spherically-symmetric collapse of a Hagedorn fluid in the brane world, the equation for outgoing, radial, null geodesics is
\begin{equation}
\frac{du}{dr} = \frac{1}{1 - \frac{2F_{H}(u)}{r} - \frac{\delta G_{H}(u) K(w)}{r}}
\end{equation}
and the corresponding expression for general relativity is obtained simply by replacing $\delta = \alpha \rho_0^2/2$ with $\rho_0/2$ and the functions $F(w)$ with $F_{GR}(w)$ and $K(w)$ with $K_{GR}(w)$, according to their respective definitions. Therefore, setting
\begin{eqnarray}
F_{H}(u) = \frac{\alpha_0 u}{2},  G_{H}(u) = \beta_0 u, 
\alpha_0 > 0,  \beta_0 > 0,
\end{eqnarray}
as in \cite{Ha03} and again assuming the limit given in Eq. (\ref{lim}), the singularity is locally naked if the equation
\begin{eqnarray} \label{quad}
\beta_0 X_0^2 - X_0 + 1 = 0,
\end{eqnarray}
admits positive, real solutions. Since the roots of  Eq. (\ref{quad}) are
\begin{eqnarray} 
X_0 =  \frac{1 \pm \sqrt{1 - 4\beta_0}}{2\beta_0},
\end{eqnarray}
for $\beta_0 < 1/4$ a naked singularity is formed, whereas $\beta_0 \geq1/4$ leads to the formation of a black hole, which mirrors the result for $q(u)=0$ obtained in \cite{Ha03}. Essentially, this result shows that the collapse of a Hagedorn fluid in the brane world does not favor the formation of black holes over naked singularities, as we may have naively expected due to the increased effective density and pressure. However, these corrections do lead, in the $\rho(u,r) \rightarrow \infty$ limit, to the creation of asymptotical flat metrics, rather than the cosmological metrics obtained in four-dimensional general relativity.

\section{Collapsing Hagedorn matter as a possible source of GRBs in the brane world} \label{Section3.5}

In \cite{Ha03}, the gravitational collapse of Hagedorn matter in the Vaidya geometry, ending in a naked singularity, was investigated  as a possible source of gamma-ray bursts (GRBs). GRBs are cosmic gamma-ray emissions with typical fluxes of the order $10^{-5}$ to $ 5 \times 10^{-4}$ erg cm${}^{-2}$ and durations from $10^{-2}$ to  $10^{3}$ s \cite{Pi92}. Since their distribution is isotropic, they are believed to have a cosmological origin and it has been suggested that they may occur at extra-galactic distances \cite{Pi92}. The widely accepted interpretation of this phenomenology is that the observable effects of GRBs are due to the dissipation of kinetic energy from a relativistically expanding compact object, though the underlying progenitor model is, as yet, unknown. Proposed models include the merger of binary neutron stars \cite{Pi92}, the capture of neutron stars by black holes \cite{Ca92}, energy emission from differentially rotating neutron stars \cite{KlRu98}, neutron star-quark star conversions \cite{ChDa96}, the gravitational collapse of rapidly rotating massive bodies, such as binaries and stellar cores, to form black holes \cite{MeRe93}, the formation of naked singularities \cite{ChJo94,Si98,HNA,Des,HaCh00a,HaIgNa00} and core-collapse supernova explosions \cite{St03}, though this interpretation remains disputed \cite{Hj03,WoBl06}. However, to date, such models have mainly been investigated within the context of general relativity \cite{Pi98}.

More recently, spurred by the advent of better observational data resulting from several experimental coalitions \cite{Zh07*,Zh10,Collab}, a plethora of alternative theories have arisen (again within general relativity), including ``cocoon emission" models \cite{ToWuMe09}, gamma-ray emission from first generation stars \cite{SuIo10},  collisional heating in a relativistically expanding jet of $e^{\pm}$ plasma \cite{Be10},
collisional heating in magnetized jets \cite{VuBePo11}, inter-collision induced magnetic reconnection and turbulence (``ICMART") models \cite{ZhYa10} and emissions from active galactic nuclei \cite{Zhetal13}. What is clear, however, is that most cosmological GRBs may be separated into two distinct classes characterized by either long durations and soft emission spectra, or short durations and hard emission spectra \cite{Na07,Tretal08,Be11}, where an observer frame time of $2$ s is usually taken as the separation line (see \cite{ZhMe04,Me06,Zh11} for reviews). Since, for long GRBs, the host galaxies are typically irregular, with intense star formation, the favored interpretation is that most (if not all) are produced during the core collapse of massive stars (collapsars). In contrast, short GRBs are usually found to originate from nearby early-type galaxies, with little star formation, which is in good agreement with the conjecture that they originate from mergers of compact binaries. However, this simple paradigm is challenged by a few unusual GRBs, so that more exotic models, such as energy emission from superconducting cosmic strings \cite{BaPaSp87+,BeHnVi01+,ChYuHa10+}, or from the collapse of ultra-high density stars described by a Hagedorn fluid \cite{Ha03}, may be viable progenitors for explaining anomalous results.

Here, we extend the investigation of naked singularities as GRB sources to the brane world scenario by considering the energy emission and timescale of collapse of a Hagedorn fluid for physically reasonable choices of $F_{H}(u)$ and $G_{H}(u)$ with appropriate parameter values. The method of analysis follows that presented in \cite{Ha03}, but is applied instead to the solutions of the brane world field equations obtained in the $\rho(u,r) \rightarrow \infty$ limit, rather than to the general-relativistic solution. For comparison to the data, we again choose the GRB observations, GRB 971214 and GRB 990123, with isotropic energy losses of order $10^{53}$ and $10^{54}$ erg, respectively \cite{Ku99}, and explicitly include factors of $c$ in our analysis.

Recall that in this limit, the mass profile $m(u,r)$ is given by Eq. (\ref{BW1}). Assuming that $dG_{H}^2(u)/du \rightarrow 0$ faster than $r$, as $r \rightarrow 0$, the time derivative of the mass at the center of the collapse, in CGS units, is
\begin{eqnarray}
\left(\frac{dm}{dt}\right)_{r=0} = \frac{c^3}{2G}\left[\frac{dF_{H}(u)}{du}\right]_{r=0}.
\end{eqnarray}
Now, if $t_{ff}$ is the time taken for a matter element at the surface of the collapsing star at $t=0$ to reach the center then, denoting $M|_{r=0} = m(t_{ff})|_{r=0}$ and assuming that $m(0)|_{r=0}=0$ , we have
\begin{eqnarray} \label{M}
M|_{r=0} = \frac{c^3}{2G} \int_{0}^{t_{ff}}\left[\frac{dF_{H}(u)}{du}\right]_{r=0} dt.
\end{eqnarray}
The integral may be evaluated approximately by using the first mean value theorem, which states that, for any function $f(t)$, $\int_{a}^{b} f(t)dt = (b-a)f(c)$ where $c \in (a,b)$ and $f(c)$ is the average value of $f(t)$ in the region $(a,b)$. Hence, in the following analysis, we will approximate the derivatives of the arbitrary function and the functions themselves by their average values over the region of integration. So, denoting
\begin{eqnarray}
F_0 =  \langle\ F_H(u)\rangle\ {}_{r=0}, \ \ \ 
\Phi_0 = \Bigg\langle\ \frac{dF_H(u)}{du}\Bigg\rangle _{r=0}, 
\end{eqnarray}
etc, Eqn. (\ref{M}) yields
\begin{eqnarray}
t_{ff} = \frac{2G}{c^3 F_0}M|_{r=0}.
\end{eqnarray}
The initial mass distribution of the Hagedorn fluid can be obtained from;
\begin{eqnarray}
\left(\frac{dm}{dr}\right)_{t=0} = \frac{c^3}{2G} \left[ \frac{dF_{H}(u)}{du} + \frac{2\delta}{r}G_{H}(u) \frac{dG_{H}(u)}{du}\right]_{t=0},
\end{eqnarray}
which, if we assume that $dG_{H}^2(u)/du \rightarrow 0$ faster than $r$, as $r \rightarrow 0$ (as mentioned above), gives
\begin{eqnarray} \label{linearm}
m(r)|_{t=0} = \frac{c^3}{2G} H_0 r,
\end{eqnarray}
where $H_0$ is the average value of the $r$-dependent function $H(r) = \left[ dF_{H}/du + (2\delta/r)G_{H}(u) dG_{H}(u)/du\right]_{t=0}$. This implies a linear profile for the initial mass distribution of the Hagedorn fluid, which is in agreement with the general-relativistic case for $q(u)=0$ \cite{Ha03}.

The difference between the brane world scenario and the corresponding situation in general relativity is that, in the former, the function $G_{H}(u)$ plays the role of an effective charge, $q_{(BW)}(u)$, at least with respect to the mass function, Eq. (\ref{BW1}), even if the physical fluid is charge neutral. Though we are of course free to set $G_{H}(u)=0$, we cannot do so independently, in the same way that we may set $q(u)=0$ without affecting the value of $G_H(u)$  when considering an uncharged Hagedorn fluid in a four-dimensional universe. However, since, for collapse on the brane, we are already forced to assume that $dG_{H}^2(u)/du \rightarrow 0$ faster than $r$, as $r \rightarrow 0$ (unless matter is converted near the core), in order to satisfy the DEC, it seems reasonable to assume an initially linear mass distribution, as in Eq. (\ref{linearm}), even in the brane world case.

Hence, if the initial radius of the collapsing region is $R$, then
\begin{eqnarray}
H_0 = \frac{2G M|_{t=0}}{c^2R},
\end{eqnarray}
where $M|_{t=0}$ is the initial total mass of the star. The change in mass at a point $r$ over a timescale $\Delta t$ may be estimated according to
\begin{eqnarray} \label{estimate}
&&\Delta \left(\frac{2G m}{c^2} \right) \approx \Bigg[ \frac{dF_{H}(u)}{du} \Delta u + \frac{2\delta}{r}G_{H}(u) \frac{dG_{H}(u)}{du}\Delta u + \nonumber\\
&&\delta G_{H}^2(u) \frac{\Delta r}{r^2}\Bigg]\approx \Bigg[ \frac{dF_{H}(u)}{du} +\frac{2\delta}{r}G_{H}(u) \frac{dG_{H}(u)}{du} + \nonumber\\
&&\frac{\delta G_{H}^2(u)}{r^r} \frac{v_f}{c}\left(1+\frac{v_f}{c}\right)^{-1}\Bigg]c\left(1+\frac{v_f}{c}\right)\Delta t,
\end{eqnarray}
where $v_f \approx \Delta r/\delta t$ is the speed of collapse measured by a local observer. Evaluating Eq. (\ref{estimate}) close to $r=0$ then gives the approximate energy release in the collapse as
\begin{eqnarray} \label{E_BW}
\frac{\Delta E_r^{BW}}{\Delta t} \approx \frac{c^5}{4G}\left[\mathcal{F}_0 + H_0 \frac{v_f}{c}\left(1+\frac{v_f}{c}\right)^{-1}\right]\left(1+\frac{v_f}{c}\right), \nonumber\\
\end{eqnarray}
where $\mathcal{F}_0 =  \left(c^3/2G\right)F_0$ is simply the average value of $F_{H}(u)$ in CGS units. This may be compared with the corresponding expression obtained in \cite{Ha03}
\begin{eqnarray} \label{E_GR}
\frac{\Delta E_r^{GR}}{\Delta t} \approx \frac{c^5}{2G}\left[\mathcal{F}_0 + 4\pi h_0 \frac{v_f}{c}\left(1+\frac{v_f}{c}\right)^{-1}\right]\left(1+\frac{v_f}{c}\right),\nonumber\\
\end{eqnarray}
where $h_0$ is the average value of the function $h(r) = \left[ dF_{H}/du + \rho_0 r dG_{H}(u)/du\right]_{t=0}$.

Therefore, although the $r$-dependent term of the mass profile for a collapsing sphere of Hagedorn fluid in the brane world is characterized by the ratio $\delta = \rho_0^2/(4\lambda_b)$, rather than $\rho_0$, as in general relativity, and, although, even for a uncharged fluid, it is proportional to $r^{-1}$ in the former (rather than $r$), which corresponds to the contribution from a nonzero charge, $q(u)$, in the latter, there is sufficient freedom in choosing the arbitrary functions $F_{H}(u)$ and $G_{H}(u)$, and hence the approximate average values of the compound functions $H(r)$ and $h(r)$, which contribute to the energy release in the brane world and general-relativistic universes, respectively, for the general forms of the expressions for $\Delta E_{r}^{BW}/\Delta t$ and $\Delta E_{r}^{GR}/\Delta t$, Eqs. (\ref{E_BW})-(\ref{E_GR}), to remain the same.

For $\Delta t  \sim 10^{-4}$, comparable to the rise time of GRBs  971214 and 990123 \cite{Ku99}, we obtain an energy emission of 
\begin{eqnarray}
\Delta E_{\gamma}^{BR} \approx \left[\mathcal{F}_0 + H_0 \frac{v_f}{c}\left(1+\frac{v_f}{c}\right)^{-1}\right] \times 10^{55} \;{\rm erg},
\end{eqnarray}
in the braneworld and
\begin{eqnarray}
\Delta E_{\gamma}^{GR} \approx \left[\mathcal{F}_0 + h_0 \frac{v_f}{c}\left(1+\frac{v_f}{c}\right)^{-1}\right] \times 10^{55} \;{\rm erg}
\end{eqnarray}
for the collapse of a Hagedorn fluid in general relativity. Clearly, since the constants $\mathcal{F}_0$, $H_0$ and $h_0$ can be tuned to be of order unity in either case, collapsing spheres of Hagedorn matter remain as viable as progenitor models for the source of gamma-ray bursts in brane world cosmology as they do in general relativity. Considering the analysis presented above from a purely from a mathematical point of view, this may not seem very surprising, given the freedom allowed in choosing the arbitrary functions. However, physically, it is important to consider two things. Firstly, to establish whether the higher order corrections to the gravitational field equations  on the brane favor the existence of black holes over naked singularities, as may be expected naively. The work presented in section III.D shows that, counter to our intuition, this is not the case. Secondly, since we may also imagine intuitively that  the increased effective density and pressure in the brane world could lead to greater energy emission in a gravitational collapse, it is important to establish whether this is necessarily true under generic conditions. The work presented in this section shows that it is not. Therefore, naked singularities as GRB sources (at least from a collapsing Hagedorn fluid) are neither favored nor disfavored by brane world cosmology as opposed to general relativity.

\section{Conclusions} \label{Section4}

We solved the approximate gravitational field equations on the brane, in the high density limit, for spherically-symmetric null fluids described by the barotropic, polytropic, strange quark ``bag" model and Hagedorn equations of state. For barotropic fluids, the brane world and general-relativistic solutions coincide only for the single parameter choice, $k=0$ on the brane (corresponding to the collapse of a null dust), and $k=1$ in general relativity, both of which give rise to the charged Vaidya metric.  While values of $k$ in the range $0 \leq k <1/2$ give rise to cosmological metrics and $1/2 < k \leq 1$ produces asymptotically flat space-times (with $k=1/2$ corresponding to the critical case in which no reasonable set of energy conditions can be fulfilled), in the general-relativistic solution, by contrast, the $\rho^2(u,r)$ corrections to the effective energy density on the brane ensure that all the space-times produced by the collapse of a barotropic fluid are asymptotically flat and no critical case exists. However, while these corrections are enough to provide asymptotic flatness, they are not sufficient to rule out the formation of locally naked singularities, which remain as viable on the brane as in the standard scenario. We also confirm the existence of a mass gap between black hole and naked singularity solutions in the brane world model, analogous to that found by Hussain in general relativity \cite{Hu96}. The metrics obtained for the collapse of a fluid governed by the polytropic equation of state on the brane also closely resemble those obtained in the standard case, but again, they are always asymptotically flat, whereas the former may be cosmological for appropriate parameter choices \cite{Hu96}.

For collapsing strange quark matter, the formation of naked singularities also remains as viable in the brane world as in general relativity and, in this case, the metrics obtained remain cosmological, as they are in the latter. In the case of strange matter collapse to form a black hole, the causal structure of the horizons (for equivalent initial conditions and collapse profiles) or, in other words, the null fluid hair possessed by the black hole, differs profoundly between the two scenarios. In principle, it is also possible for the same initial data to lead to black hole production in one scenario, but the formation of a naked singularity in the other.

Interestingly, the brane world corrections for a strange quark fluid seem to play a similar (though not identical) role to the presence of a charged fluid, governed by an arbitrary function $q(u)$, in the general-relativistic scenario. However, for the collapse of a neutral fluid, either on the brane or in general relativity, the solution is characterized by two independent functions of $u$, whereas, for a collapsing charged fluid, it  contains three (including $q(u)$). Therefore, even if the precise form of the terms in the metric introduced by the $\rho^2(u,r)$ corrections to the field equations on the brane and those containing $q(u)$ in the general-relativistic solution were identical, the collapse of an uncharged quark fluid in the brane world scenario could only be equivalent to the collapse of a charged quark fluid in general relativity if some additional constraint existed relating $q(u)$ to the other functions.

Similar results hold true for the collapse of a Hagedorn fluid, in that the existence of naked singularities again remains as viable on the brane as in general relativity, though the structure of null fluid hair in black hole solutions differs between the two models, for identical initial conditions and collapse profiles. As in the case of the quark fluid, the brane world corrections play a similar, though again not identical, role to the presence of a charged fluid component, governed by $q(u)$, but contain only two independent arbitrary functions. Unlike for strange matter collapse, however, the metrics obtained for the collapse of a Hagedorn fluid on the brane are all asymptotically flat, as opposed to cosmological in the general-relativistic case. Finally, we investigated the possibility of naked singularity formation, from the collapse of a Hagedorn fluid on the brane, as a source of GRB emissions and found that this remains as viable a progenitor model as in general relativity, being neither more nor less favored by current observational data.

\section*{Acknowledgments}
We dedicate this paper to the memory of Christopher Beling, gone but not forgotten.

\clearpage

\end{document}